\documentclass[sigconf]{acmart}

\usepackage{algorithmic}
\usepackage{graphicx}
\usepackage{textcomp}
\usepackage{xcolor}
\def\BibTeX{{\rm B\kern-.05em{\sc i\kern-.025em b}\kern-.08em
    T\kern-.1667em\lower.7ex\hbox{E}\kern-.125emX}}

\usepackage{multirow}
\usepackage{url}
\usepackage{amsmath,amsfonts}
\newcommand{\stitle}[1]{\vspace{1ex}\noindent\textbf{#1}}
\newcommand{\ignore}[1]{}

\setcopyright{none}
\renewcommand\footnotetextcopyrightpermission[1]{}
\newcounter{JohnNOC}
\stepcounter{JohnNOC}

\newcounter{VasiaNOC}
\stepcounter{VasiaNOC}

\newcommand*\rot{\rotatebox{0}}
\usepackage[font=small]{caption}
\usepackage[font=footnotesize]{subcaption}
\usepackage{fancyvrb}
\usepackage{xcolor}

\newcounter{MayankNOC}
\stepcounter{MayankNOC}

\newcommand{\mvlater}[1]{}

\usepackage{xspace}
\newcommand{\ours}{\emph{Secrecy}\xspace}
\newcommand{\opcost}{\ensuremath{C_{o}}\xspace}
\newcommand{\sync}{\ensuremath{C_{s}}\xspace}

\usepackage{subcaption}
\usepackage{graphicx}
\usepackage{tabularx}
\usepackage{booktabs}

\usepackage[vlined,linesnumbered]{algorithm2e}
\makeatletter
\renewcommand{\@algocf@capt@plain}{bottom}
\makeatother

\SetKwInOut{Input}{Input}
\SetKwInOut{vars}{vars}
\SetKwInOut{Output}{Output}
\SetKwInOut{Parameter}{Param.}
\SetKwInput{Procedure}{Procedure}
\SetKwComment{Comment}{{//}}{}
\newcommand{\Com}[1]{\Comment*{\small #1}}

\usepackage{ifthen}

\newcommand{\fullversiontext}{highlight}

\newcommand{\iffull}[2]{\ifthenelse{\equal{\fullversiontext}{off}}{#2}{\ifthenelse{\equal{\fullversiontext}{on}}{#1}{{#1}}}}

\newcommand{\boolXOR}{\ensuremath{\texttt{XOR}}\xspace}
\newcommand{\boolAND}{\ensuremath{\texttt{AND}}\xspace}
\newcommand{\rca}{\ensuremath{\texttt{RCA}}\xspace}
\newcommand{\equ}{\ensuremath{\text{eq}}\xspace}
\newcommand{\ineq}{\ensuremath{\text{ineq}}\xspace}
\newcommand{\sort}[1]{\ensuremath{\text{sort}_{#1}}\xspace}

\newcommand{\func}[1]{\ensuremath{\mathcal{F}_{\textsf{#1}}}\xspace}

\begin{document}

\title{\ours: Secure collaborative analytics on secret-shared data}

\author{John Liagouris, Vasiliki Kalavri, Muhammad Faisal, Mayank Varia}
\affiliation{Boston University} 
\affiliation{\{liagos, vkalavri, mfaisal, varia\}@bu.edu}

\begin{abstract}

We present a relational MPC framework for secure collaborative analytics on private data with no information leakage. 
Our work targets challenging use cases where data owners may not have private resources to participate in the computation, thus, they need to securely \emph{outsource} the data analysis to untrusted third parties. 
We define a set of oblivious operators, explain the secure primitives they rely on, and analyze their costs in terms of operations and inter-party communication. We show how these operators can be composed to form \emph{end-to-end oblivious queries}, and we introduce logical and physical optimizations that dramatically reduce the space and communication requirements during query execution, in some cases from quadratic to linear or from linear to logarithmic with respect to the cardinality of the input. 

We implement our framework on top of replicated secret sharing in a system called \ours and evaluate it using real queries from several MPC application areas. Our experiments demonstrate that the proposed optimizations can result in over $1000\times$ lower execution times compared to baseline approaches, enabling \ours to outperform state-of-the-art frameworks and compute MPC queries on millions of input rows with a single thread per party. 

\end{abstract}

\maketitle

\section{Introduction}\label{sec:introduction}

\mvlater{Restore references that were removed during submission}

Cryptographically secure Multi-Party Computation (MPC) enables mutually distrusting parties to perform arbitrary computations on the union of their private data while keeping the data siloed from each other (and from external adversaries) with \emph{provable security guarantees}~\cite{10.1145/3387108}.  
MPC has been deployed to protect healthcare data like disease surveillance, financial data like credit scores, advertising data like conversion rates, and more~\cite{DBLP:journals/cj/ArcherBLKNPSW18,student-taxes,DBLP:conf/fc/DamgardDNNT16,DBLP:conf/ccs/BonawitzIKMMPRS17}.

Recently, systems like Conclave~\cite{Volgushev2019Conclave},
SMCQL~\cite{Bater2017SMCQL}, SDB~\cite{wong2014secure, he2015sdb}, Senate~\cite{Poddar2021Senate}, and others~\cite{bater2020saqe, bater2018shrinkwrap} have made MPC more accessible to data analysts by providing relational interfaces and automated query planning.  To achieve practical performance,  these works employ optimizations that either leak information to untrusted parties or target peer-to-peer deployments where data owners must also serve as computing parties using private (i.e., trusted) resources.  Systems in the first class improve query performance via \emph{controlled information leakage}, e.g., by revealing intermediate result sizes, after adding noise \cite{bater2018shrinkwrap, bater2020saqe} or not \cite{Volgushev2019Conclave, wong2014secure, he2015sdb}. On the other hand, peer-to-peer systems aim to reduce joint computation to subsets of peers~\cite{Poddar2021Senate} or sidestep MPC via \emph{hybrid execution}~\cite{Bater2017SMCQL, Volgushev2019Conclave, secure-db-services, DBLP:conf/ndss/ChowLS09}, i.e., by splitting the query plan into a plaintext part (executed locally by the data owners) and an oblivious part (executed jointly~under~MPC). 

Unfortunately, existing optimizations for relational MPC are not applicable when data owners do not have private resources to participate in the protocol execution. As an example, consider a scenario where researchers from different hospitals want to conduct a large-scale medical study that requires evaluating a set of relational queries on the union of the hospital databases. The researchers cannot simply exchange the patient records, as they contain highly sensitive information about individuals and are protected by strict data privacy regulations. Even worse, the hospital IT departments might not have the domain expertise or the infrastructure needed for this type of collaborative analytics. One option is to securely outsource the computation to third parties but this raises important privacy concerns: the hospitals must ensure that no external party will be able to learn valuable personal information about patients that it could then sell to other entities in underground markets~\cite{10.1145/2068816.2068824}.

The above scenario highlights the need for outsourced relational queries on private data with no information leakage. It is representative of a family of offline analytics, where multiple data owners are willing to allow certain computations on their collective private data (e.g., for profit, social good, improved services, etc.), provided that the data remain siloed from untrusted entities. 
For example, some companies would agree to participate in a study on the gender or racial wage gap~\cite{bwwc} but only if no employee wages are revealed, as they may lose their competitive advantage. Another use case is private advertising: web users may subscribe to personalized recommendations based on collaborative filtering as long as their online activity remains hidden from the service provider.

Our work addresses the challenge of providing efficient secure collaborative analytics in the outsourced setting with no information leakage.  Outsourced MPC removes the computation burden from data owners and has recently gained attention in industry, e.g., in
 Mozilla Telemetry~\cite{prio},  Bosch's Carbyne  \cite{carbyne}, Facebook's CrypTen~\cite{crypten}
and Cape Privacy's TFEncrypted~\cite{cape_privacy}. Unfortunately, all these systems focus on simple statistics or ML workloads and do not support efficient relational analytics.  To fill this gap, we present \ours, a new relational system that allows data owners and analysts to benefit from the ``pay-as-you-go'' cost model of the cloud while retaining the \emph{full} security guarantees of MPC. 
\ours exposes the costs of oblivious queries to the planner and employs logical, physical, and protocol-specific optimizations all of which are applicable \emph{within} MPC,  even when none of the data owners participates in the computation.

\stitle{Contributions.}
We make the following contributions:\vspace{-1mm}

\begin{itemize}

\item We present a query optimization framework for outsourced relational MPC with no information leakage.
 
\item We express MPC query costs in terms of secure computation and communication operations and we propose a rich set of optimizations to improve performance: (i) database-style \emph{logical transformations}, such as operator re-ordering and decomposition, (ii) \emph{physical optimizations}, including operator fusion and message batching, and (iii) \emph{secret-sharing optimizations} that leverage knowledge about the MPC protocol.

\item We provide efficient implementations of oblivious operators and a query planner that applies the proposed optimizations. 

\item We evaluate \ours's performance and the effectiveness of the proposed optimizations using real and synthetic queries.
Our experiments show that \ours outperforms state-of-the-art MPC frameworks and scales to much larger datasets.\vspace{-1mm} 

\end{itemize}\vspace{1mm}

We will release \ours as open-source. This work aims to make MPC more accessible to the data management community and catalyze collaborations between cryptographers and database experts.

\section{\ours System Overview}\label{sec:overview}

\begin{figure}[t]
    \centering
        \begin{minipage}{.47\textwidth}
            \begin{subfigure}{\textwidth}
            \centering
  			\includegraphics[width=\linewidth]{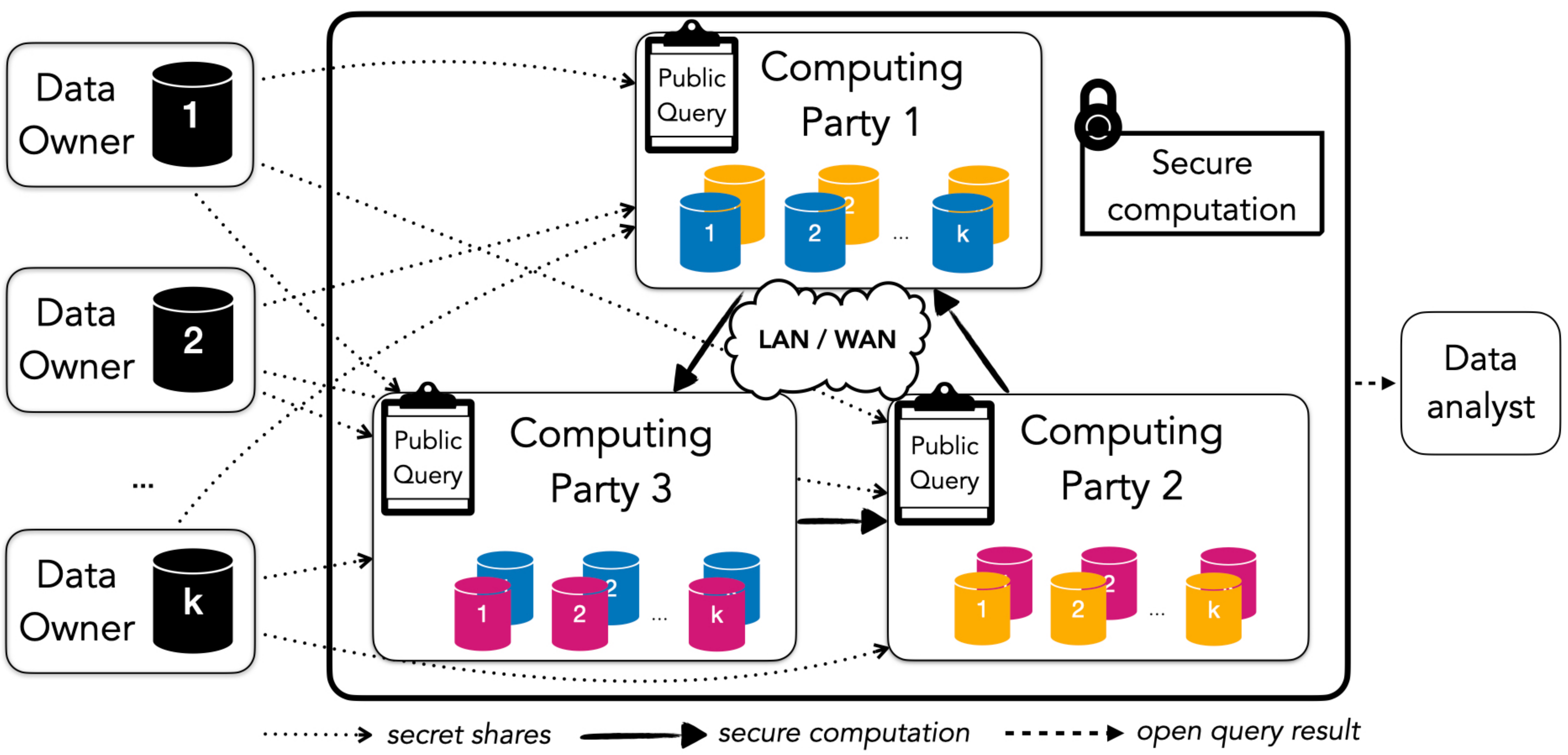}
            \end{subfigure}
        \end{minipage} 
     	 \vspace{-3mm}
        \caption{\ours allows a set of $k$ data owners to securely compute a public query on their private data by outsourcing the computation to non-colluding untrusted parties.  Data owners distribute \emph{secret shares} (cf.~\S~\ref{sec:replicated-sharing}) to the available computing parties, which execute the query under MPC and open their results to a data analyst.}\label{fig:overview}\vspace{-3mm}
    \end{figure}

Figure~\ref{fig:overview} presents an overview of the outsourced MPC setting. 
Each party in \ours has one or more of the following roles:
(i) \emph{data owner} who provides some input data,  
(ii) \emph{computing party}, e.g., a cloud provider that provides resources to perform the secure computation, and 
(iii) \emph{analyst} who issues a query to learn~the~result. 
A ``party'' is a logical entity and does not necessarily correspond to a single compute node. 
\ours does not make any assumption about the physical deployment: each party can be deployed at competing providers or within multiple providers in a federated cloud.

\subsection{Design principles}
We have designed \ours on the following principles:

\stitle{1. Decoupling of roles.}
In \ours, a party may have any combination of roles; for instance, data owners can (but do not have to) also act as computing parties and/or analysts without affecting the security guarantees.  Query optimization in \ours does not rely on data ownership and does not require data owners to participate~in~MPC.

\stitle{2. No information leakage.}
\ours reveals nothing about the data and the execution metadata to untrusted parties. It completely hides access patterns and intermediate result sizes.  \ours does not require data owners to annotate attributes as sensitive or non-sensitive, or to provide privacy budgets, and does not try to sidestep the secure computation.
It executes all query operators under MPC and protects all attributes to prevent inference attacks that exploit correlations or functional dependencies in the data. 

\stitle{3. No reliance on trusted execution environments.}
\ours does not rely on any (semi-)trusted party, honest broker or specialized secure hardware. 
To remove barriers for adoption, we target general-purpose compute and cloud. 

\stitle{4. High expressivity.} 
\ours's protocol does not pose any restriction on the types of queries that can be supported.  
While there exist many efficient protocols for specific instances of MPC operators, such as set intersection or unique-key equality joins (cf.~\S~\ref{sec:related}), these are often not composable. 
In \ours, we have decided to provide general operator implementations that are independent of the input data characteristics and can be composed with each other to create arbitrarily complex query plans.

\subsection{Security guarantees and threat model}\label{sec:guarantees}

\ours protects data throughout the entire lifecycle and treats the query itself as public. That is, \ours assumes that data owners and analysts have previously agreed on a relational query to compute and this query is known to the computing parties, as in prior works~\cite{bater2020saqe,bater2018shrinkwrap,Volgushev2019Conclave,Bater2017SMCQL,Poddar2021Senate}. To evaluate the query, computing parties execute an identical oblivious computation and exchange messages with each other according to a protocol. 

\ours relies on the semi-honest 3-party replicated secret sharing MPC protocol by Araki et al.~\cite{ArakiFLNO16,aby3}.
The protocol provides two types of security guarantees: (i) \emph{privacy}, meaning that computing parties do not learn anything about the original data, access patterns or intermediate result sizes, and (ii) \emph{correctness}, meaning that all participants are convinced that the computation output is accurate.
These guarantees hold even in the presence of an \emph{adversary} who controls $T=1$ out of the $N=3$ computing parties. 
\ours can withstand \emph{semi-honest} adversaries, i.e., adversaries that monitor the state of the party they control (e.g., by inspecting access patterns and data sent or received) without altering its execution. 
We presume that the software faithfully and securely implements the MPC protocol, that is, formal verification is out of scope for~this~work.

\subsection{MPC query optimization}

Cost-based query optimization on plaintext data relies on selectivity estimations to reduce the size of intermediate results. 
Since all operators in \ours are oblivious and do not reveal the size of intermediate data, query optimization cannot be informed by selectivity statistics.
As a consequence, traditional selectivity-based techniques for plaintext queries, such as join reordering or filter push-down, are not effective when optimizing plans under MPC. For instance, given that oblivious selections do not reduce the size of intermediate data, pushing a filter down does not improve the cost of subsequent operators in the plan. 

To provide effective optimizations under MPC, we express the cost of query plans in terms of secure computation and communication operations. In \ours, we define three types of costs:

\begin{itemize}
  \item The \textbf{operation cost}, $C_o$, which is determined by the number of primitive MPC operations per party. Primitive operations can be local ($+, \oplus$) or remote ($\times$, $\wedge$), as we explain in \S~\ref{sec:background}. 
  \item The \textbf{synchronization cost}, $C_s$, given by the number of communication rounds across parties that are inherent in MPC. A communication round corresponds to a  barrier, i.e. a synchronization point in the distributed computation, where parties must exchange data in order to proceed.
  \item The \textbf{cost of composition}, $C_c$, which is also measured in number of operations and communication rounds required to compose oblivious relational operators under MPC. 
\end{itemize}

\ours applies automatic optimizations that aim to minimize at least one of these three costs. We present a comprehensive cost analysis of oblivious relational operators and their composition in \S~\ref{sec:costs}. Contrary to plaintext query optimization where estimations are often erroneous~\cite{10.14778/2850583.2850594}, 
in  MPC we can use the typical bottom-up dynamic programming approach to compute \emph{exact} plan costs at compile time, since $C_o, C_s$, and $C_c$ do not depend on the data distribution.

\section{Background on MPC}\label{sec:background}

MPC protocols follow one of two general techniques: obscuring the truth table of each operation using Yao's \emph{garbled circuits}~\cite{yao}, or interactively performing operations over encoded data using \emph{secret sharing}~\cite{shamir}. Garbled circuits are an effective method to securely compute Boolean circuits in high-latency environments because they only need a few rounds of communication between computing parties. Secret sharing-based approaches support more data types and operators, and they consume less overall bandwidth.
In this work we employ secret sharing in the honest-majority setting
that is reasonable for many real use cases (e.g., \cite{DBLP:conf/fc/BogdanovJSV15,talviste,DBLP:journals/cj/ArcherBLKNPSW18,DBLP:conf/fc/BogetoftCDGJKNNNPST09,student-taxes}).
Looking ahead, in \S~\ref{sec:evaluation} we will demonstrate the competitiveness of secret sharing for relational queries in both LAN and WAN settings.

\subsection{(Replicated) Secret Sharing}\label{sec:replicated-sharing}

This work uses the 3-party replicated secret sharing MPC protocol by Araki et al.\ \cite{ArakiFLNO16}.
We encode an $\ell$-bit string of sensitive data $s$ by splitting it into 3 \emph{secret shares} $s_1$, $s_2$, and $s_3$ that individually have the uniform distribution over all possible $\ell$-bit strings (for privacy) and collectively suffice to specify $s$ (for correctness). Computing parties are placed on a logical ring, as shown in Figure~\ref{fig:overview}, and each party $P_i$ receives two of the shares $s_i$ and $s_{i+1}$ (i.e., $P_1$ receives $s_1$, $s_2$, $P_2$ receives $s_2$, $s_3$, and $P_3$ receives $s_3$, $s_1$). Hence, any 2 parties can reconstruct a secret, but any single party cannot.
We consider two secret sharing formats: 
\emph{boolean} secret sharing in which $s = s_1 \oplus s_2 \oplus s_3$, where $\oplus$ denotes the boolean XOR operation, and 
\emph{additive} or \emph{arithmetic} secret sharing in which $s = s_1 + s_2 + s_3 \bmod{2^\ell}$.

\subsection{Oblivious primitives}\label{sec:secure_primitives} \label{sec:cost_primitives}

In this section, we describe several oblivious primitives we use throughout our work.
These primitives allow the parties to collectively compute secret shares of many operations, without learning anything about the actual secrets.
Most of our oblivious primitives are built from boolean operations,
though some involve arithmetic operations.
We stress that our oblivious relational operators in \S~\ref{sec:costs} use these primitives in a black-box manner and it would be perfectly possible to implement primitives based on other protocols without affecting the applicability of the proposed optimizations in \S~\ref{sec:optimizations}. We further discuss the generality of \ours's optimizations in \S~\ref{sec:sec_analysis}.

\stitle{Boolean operations.}
When given boolean secret-shared data corresponding to $\ell$-bit strings $s$ and $t$, parties can compute shares of the bitwise XOR $s \oplus t$ \emph{locally} (i.e., without communication) and shares of the bitwise AND $st$ with $C_s(\boolAND) = 1$ round of communication. The operation cost is the same in both cases, i.e., $\opcost(\boolXOR) = \opcost(\boolAND) = \ell$ (we consider 1-bit boolean operations to have unit cost).

In more detail, observe that $s t = (s_1 \oplus s_2  \oplus s_3)\cdot (t_1 \oplus t_2 \oplus t_3 )$. After distributing the AND over the XOR and doing some rearrangement we have $s t = (s_1  t_1 \oplus s_1  t_2 \oplus s_2  t_1) \oplus (s_2  t_2 \oplus s_2  t_3 \oplus s_3  t_2) \oplus (s_3  t_3 \oplus s_3  t_1 \oplus s_1  t_3)$. In our replicated secret sharing scheme, each party has two shares for $s$ and two shares for $t$. More precisely, $P_1$ has $s_1, s_2, t_1, t_2$ whereas $P_2$ has $s_2, s_3, t_2, t_3$, and $P_3$ has $s_3, s_1, t_3, t_1$. Using its shares, each party can locally compute one of the three terms (in parentheses) of the last equation and this term corresponds to its boolean share of $s t$. The parties then XOR this share with a fresh sharing of 0 (which is created locally) so that the final share is uniformly distributed~\cite{ArakiFLNO16}. In the end, each party must send the computed share to its successor on the ring (clockwise) so that all parties have two shares of $s t$ without knowing the actual value of $s t$. 
Logical OR and NOT are based on the XOR and AND primitives.

\stitle{Equality/Inequality.} 
Using these boolean operations,
 parties can jointly compute $s \stackrel{?}{=} t$ (resp. $s \stackrel{?}{<} t$)
by computing a sharing of $s \oplus t$ and then taking the oblivious boolean-AND of each of the bits of this string (resp., taking the value of $s_i$ at the first bit $i$ in which the two strings differ).
As a result, taking the equality of $\ell$-bit strings requires $\opcost(\equ) = 2 \ell - 1$ operations (namely, $\ell$ XORs plus $\ell-1$ ANDs) and $\sync(\equ) = \lceil\log\ell\rceil$ rounds.
Similarly, inequality comparison has $\opcost(\ineq) = 4 \ell - 3$ 
and $\sync(\ineq) = \lceil\log(\ell+1)\rceil$.
As special cases, $s < 0$ requires no communication, and
equality with a public constant $s {=} c$ can also be done locally provided that the data owners have secret-shared the results of $s-c$ and $c-s$~\cite{crypten}. 

\stitle{Compare-and-swap.}
The parties can calculate the min and max of two strings. Setting $b = (s { < } t)$, we can use a multiplexer to compute $s' = \min\{s, t\} = b s \oplus (1 \oplus b)t$ and $t' = \max\{s, t\} = (1 \oplus b) s \oplus b t$. 
Evaluating these formulas requires 6 more operations and 1 more synchronization round beyond the cost of the oblivious inequality.

\stitle{Sort and shuffle.}
Given an array of $n$ secret-shared strings, each of length $\ell$,
oblivious sort in \ours is based on a bitonic sorter that comprises $\log n\cdot (\log n + 1) / 2$ stages and performs $n/2$ independent compare-and-swap operators in each stage.
Hence, sorting has operational cost
$\opcost(\sort{n}) = \frac{1}{4} n\log n \cdot(\log n + 1)  \cdot (\opcost(\ineq) + 6)$
and synchronization cost $\sync(\sort{n}) = \frac{1}{2} \log n \cdot(\log n + 1) \cdot (\sync(\ineq) + 1)$.
We can obliviously shuffle values in a similar fashion: each party appends an attribute that is populated with locally generated random values, sorts the values on this attribute, and then discards~it. 

\stitle{Boolean addition.}
Given boolean-shared integers $s$ and $t$, 
computing the boolean share of $s+t$
using a ripple-carry adder~\cite{DBLP:books/daglib/0014576}
can be done with $C_o(\rca) = 5\ell-3$ operations in $C_s(\rca) = \ell$ rounds.

\mvlater{The notes above about comparisons also describe an additional protocol with linear op and sync cost. We can perform the addition in $\log(\ell)$ rounds, albeit with higher $\ell \log(\ell)$ operational cost, using a parallel prefix adder \cite{aby3}.}

\stitle{Conversion.} We can convert between additive and boolean sharings \cite{aby,aby3,DBLP:journals/iacr/PatraSSY20}
by securely computing all of the XOR and AND gates in a ripple-carry adder. 
Single-bit conversion can be done in two rounds with the simple protocol used in CrypTen~\cite{crypten}. 

\stitle{Arithmetic operations.}
After converting two secrets $u$ and $v$ to additive secret sharing,
parties can compute the sum $u+v$ locally,
the product $u\cdot v$ with 1 synchronization round, and
scalar multiplication $c\cdot u$ locally given a public constant $c$.

\section{Cost analysis for Relational MPC}\label{sec:framework}\label{sec:costs}

In this section, we analyze the costs of oblivious operators in \ours (\S~\ref{sec:cost_relational}) and the costs of their composition under MPC~(\S~\ref{sec:constructing_plans}). 

\subsection{Costs of oblivious relational operators}\label{sec:cost_relational}

Let $R$, $S$, and $T$ be relations with cardinalities $|R|$, $|S|$, and $|T|$ respectively. Let also $t[a_i]$ be the value of attribute $a_i$ in tuple $t$. 
To simplify the presentation, we describe each operator based on the logical (i.e., secret) relations and not the random shares distributed across parties. That is, when we say that ``\emph{an operator is applied to a relation $R$ and defines another relation $T$}'', in practice this means that each party begins with shares of $R$, performs some MPC operations on the shares, and ends up with shares~of~$T$. 
Table~\ref{tab:costs} shows the asymptotic operation and synchronization costs per operator with respect to the input size. The detailed costs are given~below:

\stitle{PROJECT.} Oblivious projection has the same semantics as its plaintext counterpart.  The operation and synchronization costs of oblivious \texttt{PROJECT} are both zero since each party can locally disregard the shares corresponding to the filtered attributes.

\stitle{SELECT.}
An oblivious selection with predicate $\varphi$ on a relation $R$ defines a new relation:
\[
 T= \{t\cup\{\varphi(t)\}~~|~~t\in R\}\\ \\
\]
with the same cardinality as $R$, i.e. $|T|=|R|$, and one more single-bit attribute for each tuple $t\in R$ that contains $\phi$'s result when applied to $t$.  This bit denotes whether  $t$ is included in the output relation $T$ and is securely computed under MPC so that its true value remains hidden (i.e., secret-shared) from the computing parties.
Note that, in contrast to a typical selection in the clear, oblivious selection defines a relation with the same cardinality as the input, i.e., it does not remove tuples from $R$ so that the true size of $T$ is kept secret.\vspace{0.5mm} 

\noindent
\underline{\textit{Costs:}} The operation cost of \texttt{SELECT} is $C_{o}(\sigma_\phi(R)) = C_{o}(\phi(t))\cdot|R|$, \ \ $t\in R$, where $C_{o}(\phi(t))$ is the operation cost of evaluating $\phi$ on a single tuple $t\in R$. Since predicate evaluation can be performed independently for each tuple in $R$, the total number of rounds to perform the \texttt{SELECT} equals the number of rounds to evaluate the selection predicate on a single tuple, i.e., $C_s(\sigma_\phi(R)) = C_s(\phi(t)),~t\in~R$. 

Both $C_{o}(\phi(t))$ and $C_{s}(\phi(t))$ are independent of the actual $t$ contents: they only depend on $\phi$'s syntax and the lengths of the attributes used in $\phi$.
In \ours, a predicate $\phi$ can be an arbitrary logical expression with atoms that may also include arithmetic expressions ($+,\times, =, >, <, \neq, \geq, \leq$) and is constructed using the primitives of \S~\ref{sec:secure_primitives}. Consider the example predicate $\phi := \texttt{age>30 AND age<40}$ that requires \texttt{AND}ing the results of two oblivious inequalities under MPC.  Based on the costs of primitive operations from \S~\ref{sec:secure_primitives}, we have: $C_o(\phi(t)) = 2\opcost(\ineq) + \opcost(\boolAND)$ and $C_s(\phi(t)) = \sync(\ineq) + 1$.
In \S~\ref{sec:mpc_optimization}, we describe a technique we use in \ours that can reduce selections to local operations (with $C_s=0$). 

\begin{table}[t]
\scriptsize
\begin{center}
 \begin{tabular}{||c c c ||} 
 \hline
 Operator & \#operations (\#messages) & \#communication rounds \\ [0.5ex] 
 \hline\hline
\texttt{SELECT} &  $O(n)$ & $O(1)$   \\  
  \hline
\texttt{JOIN} & $O(n\cdot m)$ &$O(1)$  \\
 \hline
\texttt{SEMI-JOIN} & $O(n\cdot m)$ & $O(\log m)$  \\
 \hline
 \texttt{ORDER-BY} & $O(n\cdot \log^2 n)$ & $O(\log^2 n)$  \\
 \hline
 \texttt{DISTINCT} & $O(n\cdot \log^2 n)$ & $O(\log^2 n)$  \\ 
 \hline
 \texttt{GROUP-BY} & $O(n\cdot \log^2 n)$ & $O(\log^2n)$  \\
  \hline
  \texttt{MASK} & $O(n)$ & $O(1)$  \\
  \hline
\end{tabular}
\caption{Summary of operation and synchronization costs for general oblivious relational operators \emph{w.r.t.} the cardinalities ($n$, $m$) of the input relation(s). The asymptotic number of operations equals the asymptotic number of messages per computing party, as each individual operation on secret shares involves a constant number of message exchanges under MPC. Independent messages can be batched in rounds as shown in the rightmost column.}
 \label{tab:costs}
\end{center}\vspace{-8mm}
\end{table}

\stitle{JOIN.}  An oblivious $\theta$-join between two relations $R$ and $S$, denoted with $R\bowtie_\theta S$, defines a new relation:
\[
 T = \{(t\cup t'\cup\{\theta(t, t') \})~~|~~t\in R~\wedge~t'\in S\}\\ \\
\]
where $t\cup t'$ is a new tuple that contains all attributes of $t\in R$ along with all attributes of $t'\in S$, and $\theta(t, t')$ is $\theta$'s result when applied to the pair of tuples ($t, t'$).
$T$ is the cartesian product of the input relations ($R \times S$), where each tuple is augmented with a (secret-shared) bit denoting whether the tuple $t$ ``matches'' with tuple  $t'$ according to $\theta$.  
We emphasize that our focus in this work is on general-purpose oblivious joins that can support arbitrary predicates; there also exist special cases of oblivious join algorithms, e.g., primary- and foreign-key equi-joins with lower asymptotic complexity \cite{Krastnikov2020Efficient, database-info-sharing, psi-cuckoo-hashing, DBLP:conf/ccs/MohasselRR20} or compositions of equi-joins with specific operators \cite{cryptoeprint:2020:599} that could be added to \ours if desired (cf.~\S~\ref{sec:related}). 

\noindent
\underline{\textit{Costs:}} The general oblivious \texttt{JOIN} requires a nested-loop over the input relations to check all possible pairs, so its operation cost is $C_o(R\bowtie_\theta S) = C_o(\theta(t,t'))\cdot|R|\cdot|S|$, $t\in R, t'\in S$. However, the total number of communication rounds to evaluate the \texttt{JOIN} is independent of the input cardinality; it only depends on the join predicate $\theta$, i.e., $C_s(R\bowtie_\theta S) = C_s(\theta(t,t')), t\in R, t'\in S$. 
For example, a range join $R\bowtie_{a<b} S$ has $\opcost(R\bowtie_{a<b} S) = 2 |R|\cdot|S|\cdot\opcost(\ineq)$
and  $\sync(R\bowtie_{a<b}~S)~=~C_s(\ineq)$.
The constant asymptotic complexity in number of rounds with respect to the input cardinality holds for \emph{any} $\theta$-join. 
Join predicates in \ours can be arbitrary expressions whose cost is computed as explained  above for selection predicates.

\stitle{SEMI-JOIN.} An oblivious (left) semi-join between two relations $R$ and $S$ on a predicate $\theta$, denoted with $R\ltimes_{\theta}S$, defines a new relation:
\[
 T = \{(t\cup\{\bigvee_{\forall t'\in S}\theta(t, t')\})~~|~~t\in R\}\\ \\
\]
with the same cardinality as $R$, i.e. $|T|=|R|$, and one more attribute that stores the result of the formula 
$f(\theta, t, S) = \bigvee_{\forall t'\in S}\theta(t, t')$,\ $t\in R$
indicating whether the tuple in $R$ ``matches'' any tuple in $S$.\vspace{0.5mm}

\noindent
\underline{\textit{Costs:}} The operation cost of the general oblivious \texttt{SEMI-JOIN} is $C_o(R\ltimes_{\theta}S) = C_o(f(\theta, t, S))\cdot|R| = C_o(\theta(t,t'))\cdot|R|\cdot|S| + |R|\cdot(|S|~-~1)$, $t\in R, t'\in S$. 
The formula $f(\theta, t, S)$ can be evaluated independently for each tuple $t\in R$ using a binary tree of \texttt{OR} operations, therefore, the synchronization cost of the semi-join is $C_s(R\ltimes_{\theta} S) = C_s(\theta(t,t')) + \lceil\log|S|\rceil, t\in R, t'\in S$ (i.e., independent of $|R|$). 

\stitle{ORDER-BY.} Oblivious order-by on attribute $a_k$ has the same semantics as the non-oblivious operator. 
Hereafter, sorting a relation $R$ with $m$ attributes on ascending (resp. descending) order of an attribute $a_k, 1\leq k\leq m$, is denoted as $s_{\uparrow a_k}(R) = T$ (resp. $s_{\downarrow a_k}(R) = T$).  
We define order-by on multiple attributes using the standard semantics. For example, sorting a relation $R$ first on attribute $a_k$ (ascending) and then on $a_n$ (descending) is denoted as $s_{\uparrow a_k \downarrow a_n}(R)$.
An order-by operator is often followed by a \texttt{LIMIT} that defines the number of tuples the operator must output. 

\noindent
\underline{\textit{Costs:}} Oblivious \texttt{ORDER-BY} in \ours relies on a bitonic sorter of \S~\ref{sec:secure_primitives} that internally uses an oblivious multiplexer.
Hence, the operation and synchronization costs are $\opcost(s_{\uparrow a}(R)) = \opcost(\sort{|R|})$ and $\sync(s_{\uparrow a}(R)) = \sync(\sort{|R|})$, as given in \S~\ref{sec:secure_primitives}. 
In this case, the number of operations required by each oblivious multiplexing is linear to the number of attributes in the input relation, however, the total number of rounds depends only on the cardinality of the input. The analysis assumes one sorting attribute; adding more sorting attributes increases the number of operations and communication rounds in each comparison by a small constant factor.

\stitle{GROUP-BY with aggregation.} An oblivious group-by aggregation on a relation $R$ with $m$ attributes defines a new relation $T=\{f(t')~~|~~t'=t\cup\{a_{g}, a_v\},~~t\in R\}$ with the same cardinality as $R$, i.e. $|T|=|R|$, and two more attributes: $a_{g}$ that stores the result of the aggregation, and $a_v$ that denotes whether the tuple $t$ is `valid', i.e., included in the output.
Let $a_k$ be the group-by key and $a_w$ the attribute whose values are aggregated. Let also $S = \Big[t_1[a_w], t_2[a_w], ..., t_u[a_w]\Big]$ be the list of values for attribute $a_w$ for all tuples $t_1, t_2, ..., t_u\in R$ that belong to the same group, i.e.,   $t_1[a_k]=t_2[a_k]=...=t_u[a_k]$, $1\leq u\leq |R|$. The function $f$ in $T$'s definition above is defined as:
\[
    f(t_i)=\left\{
                \begin{array}{ll}
                  t_i[a_{g}] = agg(S), t_i[a_v] = 1,~~i=u',~~1\leq u'\leq u\\ \\
                  t_{inv},~~i\neq u',~~1\leq i\leq u\\ 
                \end{array}
              \right.
\]
where $t_{inv}$ is a tuple with $t_{inv}[a_v]=0$ and the rest of the attributes set to a special reserved value, while $agg(S)$ is the aggregation function, e.g. \texttt{MIN}, \texttt{MAX}, \texttt{COUNT}, \texttt{SUM}, \texttt{AVG}, and is implemented using the primitives of \S~\ref{sec:secure_primitives}.
Put simply, oblivious aggregation sets the value of $a_g$ for one tuple per group equal to the result of the aggregation for that group and updates (in-place) all other tuples with ``garbage.'' This operation is followed by an oblivious shuffling to hide the group boundaries when opening the relations to the learner (and only if there is no subsequent shuffling in the query plan).
Groups can be defined on multiple attributes using the standard semantics. 

\noindent
\underline{\textit{Costs:}} The \texttt{GROUP-BY} operator $\gamma^{agg}_{a_k}(R)$ breaks into two phases: an oblivious sort on the group-by key(s) and an odd-even aggregation \cite{DBLP:journals/iacr/JonssonKU11} applied to the sorted input. The odd-even aggregation performs $(|R|(\log|R| - 1) +1)\cdot C_o(agg(t,t'))$ operations in $\log|R| \cdot C_s(agg(t,t'))$ rounds, where $C_o(agg(t,t'))$ and $C_s(agg(t,t'))$ are the operation and synchronization costs, respectively, of applying the aggregation function to a single pair of tuples $t, t'\in R$ (independent of $|R|$). Accounting for the initial sorting on the group-by keys, the total operation cost of the oblivious group-by is  $C_o(\gamma^{agg}_{a_k}(R)) = C_o(s_{\uparrow a_k}(R)) + (|R|(\log|R| - 1) +1)\cdot C_o(agg(t,t'))$. The total synchronization cost is $C_s(\gamma^{agg}_{a_k}(R)) = C_s(s_{\uparrow a_k}(R)) + \log|R| \cdot C_s(agg(t,t'))$. The analysis can be easily extended to multiple group-by keys. 

\stitle{DISTINCT.} The oblivious distinct operator is a special case of group-by with aggregation, assuming that $a_k$ is not the group-by key as before but the attribute where distinct is applied. For distinct, there is no $a_g$ attribute and the function $f$ is defined as follows:
\[
    f(t_i)=\left\{
                \begin{array}{ll}
                  t_i[a_{v}]=1,~~i=u',~~1\leq u'\leq u\\ \\
                  t_i[a_{v}]=0,~~i\neq u',~~1\leq i\leq u\\
                \end{array}
              \right.
 \]
Distinct marks one tuple per group as `valid' and the rest as `invalid'.\vspace{0.5mm}

\noindent
\underline{\textit{Costs:}} The \texttt{DISTINCT} operator includes an oblivious sort on the distinct attribute(s) followed by a second phase where the operator compares adjacent tuples in the sorted input to set the distinct bit $a_v$. Setting the distinct bit for each tuple is independent from the rest of the tuples, so all distinct bit operations can be performed in bulk.
The total operation cost
$C_o(\delta{a_k}(R)) = C_o(s_{\uparrow a_k}(R)) + (|R|-1) \cdot \opcost(\equ)$
and synchronization cost
$C_s(\delta{a_k}(R)) = C_s(s_{\uparrow a_k}(R)) + \opcost(\equ)$
of oblivious distinct are dominated by the oblivious sort.
\mvlater{We can probably save on rounds by performing the equality check at the same time as the final swap gates at the last layer of the bitonic sort.}

\stitle{MASK.} Let $t_{inv}$ be a tuple with all attributes set to a special reserved value. A mask operator with predicate $p$ on a relation $R$ defines a new relation $T=\{f(t)~~|~~t\in R\}$, where $f(t) = t_{inv}*p(t) + (1-p(t))*t$,
 and $*$ denotes per-attribute multiplication. Mask is used at the end of the query, just before opening the result to the learner, and only if there is no previous masking.
 The cost analysis of \texttt{MASK} is similar to that of \texttt{SELECT} and is omitted due to lack~of~space.

\stitle{Global aggregations.}
\ours also supports global aggregations without a group-by clause. The total operation cost of a global aggregation is $C_o(agg(R)) = C_o(agg(t,t'))\cdot(|R|-1)$, where $C_o(agg(t,t'))$ is the operation cost of applying the aggregation function to a single pair of tuples $t, t'\in R$. The total synchronization cost is $C_s(agg(R)) = C_s(agg(t,t'))\cdot\lceil\log|R|\rceil$, since the aggregation can be applied using a binary tree of function evaluations.

\begin{table}[t]
\scriptsize
\begin{center}
 \begin{tabular}{||c  c ||} 
 \hline
 Operator pair(s) & \#comm. rounds \\ [0.5ex] 
 \hline\hline
  \{\texttt{SELECT},   \texttt{(SEMI-)JOIN}, \texttt{GROUP-BY}, \texttt{DISTINCT}\} $\rightarrow$ \texttt{DISTINCT} & $O(n)$  \\
  \hline
   \texttt{DISTINCT} $\rightarrow$ \{\texttt{SELECT}, \texttt{(SEMI-)JOIN}\} & $O(1)$  \\
 \hline
 \texttt{SELECT} $\leftrightarrow$ \texttt{(SEMI-)JOIN} &  $O(1)$   \\
 \hline
 \texttt{GROUP-BY} $\rightarrow$ \{\texttt{SELECT}, \texttt{(SEMI-)JOIN}\} & $O(1)$  \\
   \hline
  \{\texttt{SELECT}, \texttt{(SEMI-)JOIN}, \texttt{DISTINCT}, \texttt{GROUP-BY}\}  $\rightarrow$ \texttt{GROUP-BY}  & $O(\log^2n)$ \\ 
 \hline
\end{tabular}
\caption{Summary of composition costs in number of rounds for pairs of oblivious operators in \ours \emph{w.r.t} the number of generated tuples ($n$). Arrows denote the order of applying the two operators. Composition incurs a small constant number of boolean operations per tuple, so the cost in number of operations is $O(n)$ for all pairs.}
 \label{tab:composition}\vspace{-8mm}
\end{center}
\end{table}

\subsection{Cost of composing oblivious operators}\label{sec:constructing_plans}

Consider the composition of two operators defined as applying the second operator to the output of the first operator.
One merit of our approach is that all operators of \S~\ref{sec:cost_relational} reveal nothing about their output or access patterns, so they can be arbitrarily composed into an \emph{end-to-end oblivious query} plan without special treatment. 

Let $op_1$ and $op_2$ be two oblivious operators. In general, the composition $op_2(op_1(R))$ has an \emph{extra cost} (additional to the cost of applying the operators $op_1$ and $op_2$) because it requires evaluating under MPC a logical expression $e_c$ for each generated tuple. We define the \emph{composition cost} of $op_2(op_1(R))$ as the cost of evaluating $e_c$ on all tuples generated by $op_2$. The expression $e_c$ depends on the types of operators, as described below. Table~\ref{tab:composition} summarizes the composition costs for different pairs of \ours operators. 
    
\stitle{Composing selections and joins.} 
Recall that selections, joins, and semi-joins append a single-bit attribute to their input relation that indicates whether the tuple is included in the output.
To compose a pair of such operators, we compute both single-bit attributes and take their conjunction under MPC.
For example, for two selection operators $\sigma_{1}$ and $\sigma_{2}$ with predicates $\varphi_{1}$, $\varphi_{2}$, the composition $\sigma_{2}(\sigma_{1}(R))$ defines a new relation $T = \{t \cup \{e_c = \varphi_{1}(t) \wedge \varphi_{2}(t)\}~~|~~t \in R\}$.
The cost of composition in this case is the cost of evaluating the expression $\varphi_{1}(t) \wedge \varphi_{2}(t)$ for each tuple in $T$. This includes $|T|$ independent boolean ANDs which can be evaluated in one round. 

\stitle{Composing distinct with other operators.}
Applying a selection or a (semi-)join to the result of \texttt{DISTINCT} requires one communication round to compute the conjunction of the selection or (semi-) join bit with the bit $a_v$ generated by distinct.
However, applying \texttt{DISTINCT} to the output of a selection, a (semi-)join or a group-by operator, requires some care.
Consider the case where \texttt{DISTINCT} is applied to the output of a selection. Let $a_\phi$ be the attribute added by the  selection and $a_k$ be the distinct attribute. To set the distinct bit $a_v$ at each tuple, we must make sure there are no other tuples with the same attribute $a_k$, with $a_\phi = 1$, and whose distinct bit $a_v$ is already set. 
More formally:
\[
    t_i[a_v]=\left\{
                \begin{array}{ll}
                  1,~~\textnormal{iff}~\nexists t_j, i\neq j:t_i[a_k]=t_j[a_k] \wedge t_j[a_\phi]=1 \wedge t_j[a_v]=1  \\ \\
                  0,~~\textnormal{otherwise}\\
                \end{array}
              \right.
 \]
To evaluate the above formula, the distinct operator must process tuples \emph{sequentially}  and the composition itself requires $n$ rounds, where $n$ is the cardinality of the input. This results in a significant increase over the $O(\log^2n)$ rounds required by distinct when applied to a base relation. Applying distinct to the output of a group-by or (semi-)join incurs a linear number of rounds for the same reason. In \S~\ref{sec:physical}, we propose an optimization that reduces the cost of these compositions to a logarithmic~factor. 

\stitle{Composing group-by with other operators.}
To perform a group-by on the result of a selection or (semi-)join, the group-by operator must apply the aggregation function to all tuples in the same group that are also included in the output of the previous operator.
Consider the case of applying group-by to a selection result. To identify the aforementioned tuples, we need to evaluate the formula:
\[
b \leftarrow b \wedge t_i[a_\phi] \wedge t_{j}[a_\phi]
\]
at each step of the group-by operator, where $b$ is the bit that denotes whether the tuples $t_i$ and $t_{j}$ belong to the same group and $a_\phi$ is the selection bit. This formula includes two boolean ANDs that require two communication rounds. Applying group-by to the output of a (semi-)join has the same composition cost; in this case, we replace $a_\phi$ in the above formula with the (semi-)join attribute $a_\theta$.

To apply a selection to the result of \texttt{GROUP-BY}, we must compute a boolean AND between the selection bit $a_\phi$ and the `valid' bit $a_v$ of each tuple generated by the group-by. The cost of composition in number of rounds is independent of the group-by result cardinality, as all boolean ANDs can be applied in bulk. The same holds when applying a (semi-)join to the output of group-by. 
Finally, composing two group-by operators has the same cost with applying \texttt{GROUP-BY} to the result of selection, as described above. 

\stitle{Composing order-by with other operators.} Composing \texttt{ORDER-} \texttt{BY} with other operators is straight-forward. 
Applying an operator to the output of order-by has zero composition cost. The converse operation,
applying \texttt{ORDER-BY} to the output of an operator, requires a few more boolean operations per oblivious compare-and-swap (due to the attribute/s appended by the previous operator), but does not incur additional communication rounds.

\section{MPC Query Optimization}\label{sec:optimizations}

In this section, we present the set of \ours optimizations for efficient outsourced MPC: (i) logical transformation rules, such as operator reordering and decomposition (\S~\ref{sec:logical}), (ii) physical optimizations, such as message batching and operator fusion (\S~\ref{sec:physical}), and (iii) secret-sharing optimizations that further reduce the number of communication rounds for certain operators (\S~\ref{sec:mpc_optimization}). 

\stitle{Target queries.} Our work focuses on collaborative analytics under MPC where two or more data owners want to outsource queries on their collective data without compromising privacy. We consider all inputs as sensitive and assume that data owners wish to protect their raw data and avoid revealing attributes of base relations in query results. For example, employing MPC to compute a query that includes patient names along with their diagnoses in the \texttt{SELECT} clause is pointless. Thus, we target queries that return global or per-group aggregates and/or distinct results, as in prior works.

\stitle{Optimization rationale.} The optimizations we propose are based on the cost analysis of \S~\ref{sec:costs} and the following observations: 

\begin{enumerate}

\item With the exception of order-by with LIMIT, oblivious operators never reduce intermediate data. 

\item Oblivious join is the only operator that produces an output larger than its input.

\item The synchronization cost of blocking operators depends on the input cardinality.

\item When distinct follows a selection, a join or a group-by, the cost of composition increases from a constant to a linear number of rounds (cf. Table~\ref{tab:composition}). 

\item Many messages during MPC execution are independent.

\end{enumerate}

Guided by (1)-(3), we propose transformation rules that reduce the operation and synchronization costs (\S~\ref{sec:logical}). Guided by (4), we propose optimizations that reduce the composition cost (Sec. \S~\ref{sec:distinct-fusion}). Guided by (5), we implement message batching (\S~\ref{sec:batching}) to amortize network I/O. The rest of the optimizations (\S~\ref{sec:mpc_optimization}) leverage knowledge about the MPC protocol. 

\subsection{Computing plan costs}\label{sec:plan-costs}
\ours's query planner is based on a typical bottom-up dynamic programming algorithm that computes exact plan costs based on the analytical cost model of \S~\ref{sec:costs}. Each time an operator $op$ is added to a plan, \ours computes the operation and synchronization costs $C_o(op)$ and $C_s(op)$ using the formulas from \S~\ref{sec:cost_relational}. If the operator is applied to the output of another operator, e.g., $op_i(op_j(..)),~$\ours also computes the composition cost $C_c$. To do so, it augments the current plan with a special operator $op_{e_c}(op_i(op_j(..)))$ that applies the composition predicate $e_c$, as explained in \S~\ref{sec:constructing_plans}. $C_o(op_{e_c})$ and $C_s(op_{e_c})$ amount to the cost of composing the operators $op_i$ and $op_j$ in number of operations and rounds respectively. 
For a query with $k$ operators, the total cost is computed as $\sum_{i=1}^{k} \alpha C_o(i) + \beta C_s(i)$, where $\alpha, \beta$ are parameters of the deployment. 

\ours also treats each phase of the \texttt{GROUP-BY} and \texttt{DISTINCT} (cf.~\S~\ref{sec:cost_relational}) as a separate operator. That is, each group-by and distinct operator is split into a sorting operator (which is the most expensive) followed by a second operator that applies the odd-even aggregation (for group-by) or the equality checks (for distinct). 

\subsection{Logical transformation rules}\label{sec:logical}
We propose three types of logical transformation rules that reorder and decompose pairs of operators to reduce the MPC  costs:

\subsubsection{Blocking operator push-down}\label{sec:blocking-pushdown}

Blocking oblivious operators (\texttt{GROUP-BY}, \texttt{DISTINCT}, \texttt{ORDER-BY}) materialize and sort their entire input before producing any output tuple. 
Contrary to a plaintext optimizer that would most likely place sorting after selective operators, in MPC we have an incentive to push blocking operators down, as close to the input as possible. Since oblivious operators do not reduce the size of intermediate data, sorting the input is clearly the best option. 
Blocking operator push-down 
can provide significant performance improvements in practice, even if the asymptotic costs do not change. As an example, consider the rule that pushes \texttt{ORDER-BY} before a selection, i.e., $s_{\uparrow a}(\sigma_\phi (R)) \rightarrow \sigma_\phi(s_{\uparrow a}(R))$. Although this rule would not generate a more efficient plan in plaintext evaluation, it does so in the MPC setting. Recall that the operations required by the oblivious \texttt{ORDER-BY} depend on the cardinality and the number of attributes of the input relation (cf.~\S~\ref{sec:cost_relational}). Applying the selection after the order-by reduces the actual (but not the asymptotic) operation cost, as $\sigma_\phi$ appends one attribute~to~$R$. 

\stitle{Applicability.}  Rules in this class are valid relational algebra transformations with no special applicability conditions~under~MPC.

\subsubsection{Join push-up}\label{sec:join-pushup}

The second class of rules leverage the fact that \texttt{JOIN} is the only operator whose output is larger than its input. Based on this, we have an incentive to perform joins as late as possible in the query plan so that we avoid applying other operators to join results, especially those that require materializing the join output. For example, placing a blocking operator after a join requires sorting the cartesian product of the input relations, which increases the operation cost of the blocking operator to $O(n^2\log^2 n)$ and the synchronization cost~by~$4\times$.

\stitle{Example.} Consider the following query:\vspace{1mm} 

{\small
\textbf{Q1:}~~\texttt{SELECT DISTINCT R.id}

\hspace{7.5mm}\texttt{FROM R, S}

\hspace{7.5mm}\texttt{WHERE R.id = S.id}\vspace{1mm}
}

\noindent 
and the rule $\delta_{id}(R\bowtie_{id=id}S) \rightarrow \delta_{id}(R)\bowtie_{id=id}\delta_{id}(S))$.
Let $R$ and $S$ have the same cardinality $n$.
A plan that applies \texttt{DISTINCT} after the join operator
requires $O(n^2 \log^2 n)$ operations. On the other hand, pushing \texttt{DISTINCT} before \texttt{JOIN} reduces the operation cost to $O(n^2)$ and the composition cost from $O(n^2)$ to $O(1)$ in number of rounds.
The asymptotic synchronization cost is the same for both plans, i.e. $O(\log^2 n)$, but the actual number of rounds when \texttt{DISTINCT} is pushed before \texttt{JOIN}~is~$4\times$~lower.  

\stitle{Applicability.} Rules in this class have the same applicability conditions as similar rules for plaintext queries~\cite{surajit, larson}, even though their goal is different. In our setting, the re-orderings do not aim to reduce the size of intermediate data. In fact, a plan that applies \texttt{DISTINCT} on a \texttt{JOIN} input produces exactly the same amount of intermediate data as a plan where \texttt{DISTINCT} is placed after \texttt{JOIN}, yet our analysis reveals that the second plan has higher MPC costs. 

 \begin{algorithm}[t]
\small

sort input relation $R$ on $a_\theta$, $a_k$; 

let $d \leftarrow |R| / 2$\Com{Distance of tuples to aggregate}

\While {$d\geq1$} {
 \For {each pair of adjacent tuples ($t_i$, $t_{i+d}$), $0\leq i < |R| - d$,} {
  
  \Comment{Are tuples in the same group?}
  
  let $b \leftarrow t_i[a_k] \stackrel{?}{=} t_{i+d}[a_k]$;
  
   \Comment{Are tuples in the semi-join output too?}
    
  let $b_c\leftarrow b~\wedge~t_i[a_\theta]~\wedge~t_{i+d}[a_\theta]$\Com{$b_c$ is a bit}
   
  \Comment{Aggregation}
  
  $t_{i}[a_g] \leftarrow b_c\cdot\Big(t_{i}[a_g] + t_{i+d}[a_g]\Big) + (1-b_c)\cdot t_{i}[a_g]$;

  $t_{i+d}[a_v] \leftarrow \lnot b_c$\Com{$a_v$ is the `valid' bit}
   
  \Comment{Masking}
  $t_{i+d} \leftarrow b_c* t_{inv}~~+~~(1-b_c)* t_{i+d}$;  
  
 }
  $d = d / 2$;
}
mask remaining tuples with $t[a_v] = 0$ and shuffle $R$; 

\caption{\small $2^{nd}$ phase of Join-Aggregation decomposition.}\label{alg:composite_group2}
\end{algorithm}

\subsubsection{Join-Aggregation decomposition}\label{sec:decomposition}
Consider a query plan where a \texttt{JOIN} on attribute $a_j$ is followed by a \texttt{GROUP-BY} on another attribute $a_k\neq a_j$. 
In this case, pushing the \texttt{GROUP-BY} down does not produce a semantically equivalent plan. 
Still, we can optimize the plan by decomposing the aggregation in two phases and push the first and most expensive phase of \texttt{GROUP-BY} before the~\texttt{JOIN}. 

Let $R$, $S$ be the join inputs, where $R$ includes the group-by key $a_k$.
The first phase of the decomposition sorts $R$ on $a_k$ and computes a semi-join (\texttt{IN}) on $a_j$, which appends two attributes to $R$: an attribute $a_\theta$ introduced by the semi-join, and a second attribute $a_{g}$ introduced by the group-by (cf.~\S~\ref{sec:cost_relational})\footnote{In case the aggregation function is \texttt{AVG}, we need to keep the value sum (numerator) and count (denominator) as separate secret-shared attributes in $R$.}. During this step, $a_g$ is initialized with a partial aggregate for each tuple in $R$ (we come back to this later).

In the second phase, we compute the final aggregates per $a_k$ using Algorithm~\ref{alg:composite_group2}, which takes into account the attribute $a_\theta$ and updates the partial aggregates $a_g$ in-place using odd-even aggregation. The decomposition essentially replaces the join with a semi-join and a partial aggregation in order to avoid performing the aggregation on the cartesian product $R\times S$. This way, we significantly reduce the number of operations and communication rounds, but also ensure that the space requirements remain bounded by $|R|$ since the join output is not materialized. Note that this optimization is fundamentally different than performing a partial aggregation in the clear (by the data owners) and then computing the global aggregates under MPC \cite{Bater2017SMCQL, Poddar2021Senate}; in our case, all data are secret-shared amongst parties and \emph{both} phases are under MPC.

\stitle{Example.} Consider the following query:\vspace{1mm}

{\small
\textbf{Q2:}~~\texttt{SELECT R.$a_k$, COUNT(*)}

\hspace{7.5mm}\texttt{FROM R, S}

\hspace{7.5mm}\texttt{WHERE R.id = S.id}

\hspace{7.5mm}\texttt{GROUP BY R.$a_k$}\vspace{1mm}
}

Let $R$ and $S$ have the same cardinality $n$. 
The plan that applies \texttt{GROUP-BY} to the join output 
requires $O(n^2\log^2 n)$ operations and $O(\log^2n)$ communication rounds. When decomposing the aggregation in two phases, the operation cost is reduced to
 $O(n^2)$ and the synchronization cost is $4\times$ lower. The space requirements are also reduced from $O(n^2)$ to  $O(n)$. In our example, the partial aggregation corresponds to the function $t[a_\theta] = \gamma'(\theta, t, S) = \sum_{\forall t'\in S}{\theta(t,t')}, t\in R$, where $\theta(t,t') := t[\texttt{id}]\stackrel{?}{=}t'[\texttt{id}]$. This function replaces the default semi-join formula from \S~\ref{sec:cost_relational}. Similar partial aggregations can be defined for \texttt{SUM}, \texttt{MIN/MAX},~and~\texttt{AVG}.
 
\stitle{Decomposition with DISTINCT.} Decomposition can also be used when the join is followed by a \texttt{DISTINCT} operator to push the sorting phase of distinct before the join. The decomposition rule in this case is $\delta_a(R\bowtie_{\theta} S) \rightarrow \delta'_a(s_{\uparrow a}(R) \ltimes_{\theta} S)$, where $\delta'(\cdot)$ denotes the second phase of distinct that computes the distinct bit by checking adjacent tuples (cf.\S~\ref{sec:cost_relational}). 
For example, the plan $\delta_a(R\bowtie_{a=a~\texttt{AND}~a=b} S)$ can be replaced with the equivalent plan $\delta'_a(s_{\uparrow a}(R)\ltimes_{a=a~\texttt{AND}~a=b} S)$ to reduce the operation cost from $O(n^2\log^2n)$ to $O(n^2)$ and the synchronization cost from $O(n^2)$ to $O(\log^2n)$. 

\stitle{Applicability.} The decomposition technique we described is applicable to any $\theta$-join followed by (i) a \texttt{GROUP-BY} with aggregation or (ii) a \texttt{DISTINCT} operator, under the condition that the group-by or distinct keys belong to one of the join inputs. 

\subsection{Physical optimizations}\label{sec:physical}
We now describe a set of physical optimizations in \ours.

\subsubsection{Predicate fusion}\label{sec:predicate-fusion}
Fusion is a common optimization in plaintext query planning, where the predicates of multiple filters can be merged and executed by a single operator. Fusion is also applicable to oblivious selections and joins with equality predicates, and is essentially reduced to identifying independent operations that can be executed within the same communication round.
For example, if the equality check of an equi-join and a selection are independent of each other, a fused operator requires $\lceil\log\ell\rceil + 1$ rounds instead of $2 \lceil\log\ell\rceil + 1$. Next, we describe a somewhat more interesting fusion.

\subsubsection{Distinct fusion}\label{sec:distinct-fusion}
Recall that applying \texttt{DISTINCT} after \texttt{SELECT} requires $n$ communication rounds (cf.~\S~\ref{sec:constructing_plans}). We can avoid this overhead by fusing the two operators in a different way, that is, sorting the input relation on the selection bit first and then on the distinct attribute. Sorting on two (instead of one) attributes adds a small constant factor to each oblivious compare-and-swap operation, hence, the asymptotic complexity of the sorting step remains the same. When distinct is applied to the output of other operators, including selections and (semi-)joins, this physical optimization keeps the number of rounds required for the composition low. 

\stitle{Example.} Consider the following query:\vspace{1mm}

{\small
\textbf{Q3:}~~\texttt{SELECT DISTINCT id}

\hspace{7.5mm}\texttt{FROM R}

\hspace{7.5mm}\texttt{WHERE  $a_k$ = `c'}\vspace{1mm}
}

Fusing the distinct and selection operators reduces the number of communication rounds from $O(n)$ to $O(\log^2 n)$, as if the distinct operator was applied only to $R$ (without a selection). \texttt{DISTINCT} can be fused with a join or a semi-join operator in a similar way. In this case, the distinct operator takes into account the equality or inequality predicate of the (semi-)join.

\subsubsection{Message batching}\label{sec:batching}
In communication-intensive MPC tasks, some non-local operations require exchanging very small messages. Grouping and exchanging small \emph{independent} messages in bulk improves performance significantly. Consider applying a selection with an equality predicate on a relation with $n$ tuples. Performing oblivious equality on one tuple requires $\lceil\log\ell\rceil$ rounds (cf.~\S~\ref{sec:secure_primitives}). Applying the selection tuple-by-tuple and sending messages eagerly (as soon as they are generated) results in $n \cdot \lceil\log\ell\rceil$ communication rounds. Instead, if we apply independent selections across the entire relation and exchange messages in bulk, we can reduce the total synchronization cost to $\lceil\log\ell\rceil$. We apply this optimization by default to all oblivious operators in \ours. Costs in Tables~\ref{tab:costs} and~\ref{tab:composition} already take message batching into account. In~\S~\ref{sec:evaluation}, we show how message batching amortizes the (otherwise prohibitive) communication cost of secret-sharing protocol in the WAN setting. 
 
\subsection{Secret-sharing optimizations}\label{sec:mpc_optimization}

In this final subsection, we propose optimizations that take advantage of mixed-mode MPC protocols that permit both arithmetic and boolean computations.
While \ours uses boolean secret sharing for most operations, computing arithmetic expressions or aggregations like \texttt{COUNT} and \texttt{SUM} on boolean shares requires using a ripple-carry adder (\texttt{RCA}), which in turn requires inter-party communication.
Performing these operations on additive shares would require no communication, but converting shares from one format to another can be expensive.
Below, we describe two optimizations that avoid the \texttt{RCA} in aggregations and predicates with constants.

\subsubsection{Dual sharing}\label{sec:dual-sharing}
The straight-forward approach of switching from boolean to additive shares (and vice versa) based on the type of operation does not pay off; the conversion itself relies on the ripple-carry adder (cf.~\S~\ref{sec:secure_primitives}), which has to be applied twice to switch to the other representation and back. 
The cost-effective way would be to evaluate logical expressions using boolean shares and arithmetic expressions using additive shares. However, this is not always possible because arithmetic and boolean expressions in oblivious queries often need to be composed into the same formula. We mitigate this problem using a dual secret-sharing scheme.

Recall the example query \textbf{Q2} from \S~\ref{sec:decomposition} that applies an aggregation function to the output of a join according to Algorithm~\ref{alg:composite_group2}. The attribute $a_\theta$ in Algorithm~\ref{alg:composite_group2} is a single-bit attribute denoting that the respective tuple is included in the join result. During oblivious evaluation, each party has a boolean share of this bit that is used to compute the arithmetic expression in line \textbf{6}. The na\"ive approach is to evaluate the following equivalent logical expression directly on the boolean shares of $b_c$, $t_i[a_g]$, and $t_{i+d}[a_g]$:\vspace{-2mm}

\[
 t_{i}[a_g] \leftarrow b_\ell\wedge\texttt{RCA}\Big(t_{i}[a_g], t_{i+d}[a_g]\Big)~~\oplus~~ \overline{b}_\ell\wedge t_{i}[a_g]
\]

\noindent
where \texttt{RCA} is the oblivious ripple-carry adder primitive, $b_\ell$ is a string of $\ell$ bits (the length of $a_g$) all of which are set equal to $b_c$, and $\overline{b}_\ell$ is the binary complement of $b_\ell$.
Evaluating the above expression requires $\ell$ communication rounds for \texttt{RCA} plus two more rounds for the logical ANDs ($\wedge$). On the contrary, \ours evaluates the equivalent formula in line \textbf{6} of Algorithm~\ref{alg:composite_group2} in four rounds (independent from $\ell$) as follows. First, parties use arithmetic shares for the attribute $a_g$ to compute the addition locally. Second, each time they compute the bit $b_c$ in line \textbf{5}, they exchange boolean as well as arithmetic shares of its value. To do this efficiently, we rely on the single-bit conversion protocol used also in CrypTen~\cite{crypten}, which only requires two rounds of communication. Having boolean and arithmetic shares of $b_c$ allows us to use it in boolean and arithmetic expressions without paying the cost of \texttt{RCA}. 

\subsubsection{Proactive sharing}\label{sec:proactive-sharing} 
The previous optimization relies on $b_c$ being a single bit. In many cases, however, we need to compose boolean and additive shares of arbitrary values.
Representative examples are join predicates with arithmetic expressions on boolean shares, e.g. $(R.a-S.a\geq c)$, where $a$ is an attribute and $c$ is a constant. We can speedup the oblivious evaluation of such predicates by proactively asking the data owners to send shares of the expression results. In the previous example, if parties receive boolean shares of  $S.a + c$ they can avoid computing the boolean addition with the ripple-carry adder. A similar technique is also applicable for selection predicates with constants. In this case, to compute $a > c$, if parties receive shares of $a-c$ and $c-a$, they can transform the binary equality to a local comparison with zero (cf.~\S~\ref{sec:secure_primitives}).
Note that proactive sharing is fundamentally different than having data owners perform local filters or pre-aggregations prior to sharing. In the latter case, the computing parties might learn the selectivity of a filter or the number of groups in an aggregation (if results are not padded). In our case, parties simply receive additional shares and will not learn anything about the intermediate query results.

\section{Security Analysis}\label{sec:sec_analysis}

We have purposely designed \ours in a modular \emph{black-box} fashion, with a hierarchy of: MPC protocol functionalities $\to$ oblivious primitives $\to$ relational operators $\to$ optimizations.
This design choice provides two benefits: (i) immediate \emph{inheritance} of all security guarantees provided by the underlying MPC protocol, and (ii) \emph{flexibility} to support different protocols in the future that might have a different number of parties, threshold, and threat model.

\stitle{Inheritance of security guarantees.}
\ours relies on a set of functionalities that must be provided by the MPC protocol. These functionalities enable parties to receive secret-shared inputs and return secret-shared outputs: 
(i) $\func{add}$ and $\func{mult}$ that add and multiply their inputs, (ii) $\func{xor}$ and $\func{and}$ that take boolean operations of their inputs, (iii) $\func{a2b}$ and $\func{b2a}$ that perform conversions between arithmetic and boolean representations, (iv) $\func{eq}$ and $\func{cmp}$ to compute the equality and comparison predicates (where the hardest step of the latter usually involves extracting the most significant bit of an arithmetic-shared value), and (v) $\func{sh}$ and $\func{rec}$ that allow external participants to secret-share data to and reconstruct data from the computing parties.

In this section, we argue that \ours retains the security guarantees provided by the underlying MPC protocol, or equivalently that it retains the security guarantees of these ideal functionalities.
Our reasoning shows that \ours compiles each query into a sequence of calls to these functionalities that is \emph{oblivious}, meaning that its control flow is independent of its input and all data remains hidden:

\begin{enumerate}

\item \ours calls the functionalities of the MPC protocol in a black-box manner. As a result, computing parties always operate on secret-shared data; only $\func{rec}$ provides any data in the clear (namely to the learner), and \ours only calls this functionality once at the end of the query execution.

\item The control flow of each relational operator (\S~\ref{sec:cost_relational}) is oblivious, i.e., data-independent. Concretely, \texttt{SELECT} and \texttt{PROJECT} always require a single pass over the input, (semi-)\texttt{JOIN}s require a nested for-loop over the two inputs, \texttt{ORDER-BY} is based on an oblivious sorting network, and \texttt{GROUP-BY} and \texttt{DISTINCT} consist of an \texttt{ORDER-BY} followed by an additional oblivious step (to apply the aggregation and identify the unique records, respectively).

\item \ours composes relational operators (\S~\ref{sec:constructing_plans}) using the protocol functionalities (e.g., taking ANDs under MPC) within an oblivious linear scan over the output of the composition. 

\item The logical and physical transformations (\S~\ref{sec:optimizations}) rewrite the oblivious sequence of calls to the protocol functionalities into a new semantically equivalent sequence of calls that is also oblivious and has lower execution cost. 

\end{enumerate}

As a result, semi-honest security of the full \ours protocol follows by inspection of the ideal functionalities.
Privacy is satisfied against all parties because none of the functionalities ever provides a (non-secret-shared) output to the data owners or computing parties, and only the final $\func{rec}$ provides an output to the analyst as desired.
Correctness of the full protocol follows immediately from correctness of each individual functionality.

\stitle{Generality of optimizations.}
The logical and physical query optimizations constructed in this work (\S~\ref{sec:logical}-\ref{sec:physical}) apply generally to any mixed-mode MPC protocol that supports the
set of functionalities we describe above. 
This level of abstraction is commonly used by modern mixed-mode MPC protocols 
(e.g., \cite{aby,aby3,ArakiFLNO16,DBLP:conf/ccs/ChaudhariCPS19,DBLP:conf/ndss/PatraS20,DBLP:conf/ndss/ChaudhariRS20,DBLP:journals/iacr/PatraSSY20,DBLP:conf/uss/KotiPPS21,DBLP:conf/uss/Dalskov0K21}).

If providing malicious security, we require these functionalities to validate the shares of their inputs and outputs (e.g., using an information-theoretic MAC or replicated sharing), either immediately or with delayed validation before invoking $\func{rec}$.
As a consequence, \ours satisfies correctness against the computing parties because input validation binds them to provide the output of the prior step as the input shares into the next functionality.
Additionally, correctness against the data owners and analyst follow from the fact that, aside from the data owners' initial sharing through $\func{sh}$, none of the functionalities allow them to provide an input so they cannot influence the protocol execution.

As a result, the techniques from \ours can be applied to any $N$-party MPC protocol that provides semi-honest or malicious security against $T$ adversarial parties.
In particular, \ours can be instantiated with 2, 3, and 4-party secret sharing-based protocols that remain secure in the face of a malicious adversary who can deviate from the protocol arbitrarily (e.g., \cite{DBLP:journals/iacr/PatraSSY20,DBLP:conf/uss/KotiPPS21,DBLP:conf/uss/Dalskov0K21}), or with 
(authenticated) garbled circuit protocols \cite{yao,DBLP:conf/ccs/WangRK17} combined with occasional conversions to arithmetic secret sharing \cite{aby,DBLP:journals/iacr/PatraSSY20} as needed.
Protocols that provide the stronger cryptographic guarantee of robustness often do so by running several MPC executions both before and after evicting the malicious party, and by the same logic as above \ours even maintains the robust security of these protocols.

In contrast to the above, we remark that the secret-sharing optimizations of \S~\ref{sec:mpc_optimization} are specific to
the Araki et al.~\cite{ArakiFLNO16} protocol used within \ours (cf.~\S~\ref{sec:replicated-sharing}). That said, we expect that similar techniques to reduce the number of operations and/or communication rounds can be developed for other protocols.

\stitle{Alternative oblivious primitives.} \ours can also support alternative instantiations of individual oblivious primitives with different cost profiles, such as constant-round equality and comparisons with higher operational costs \cite{DBLP:conf/tcc/DamgardFKNT06,DBLP:journals/ieicet/NishideO07}.
Extending our query planner to consider the cost profiles offered by a variety of building blocks is an exciting opportunity for future work (cf. \S~\ref{sec:next}).

\section{\ours Implementation}\label{sec:implementation}

  \begin{figure}[t]
    \centering
        \begin{minipage}{.47\textwidth}
            \centering
  			\includegraphics[width=\linewidth]{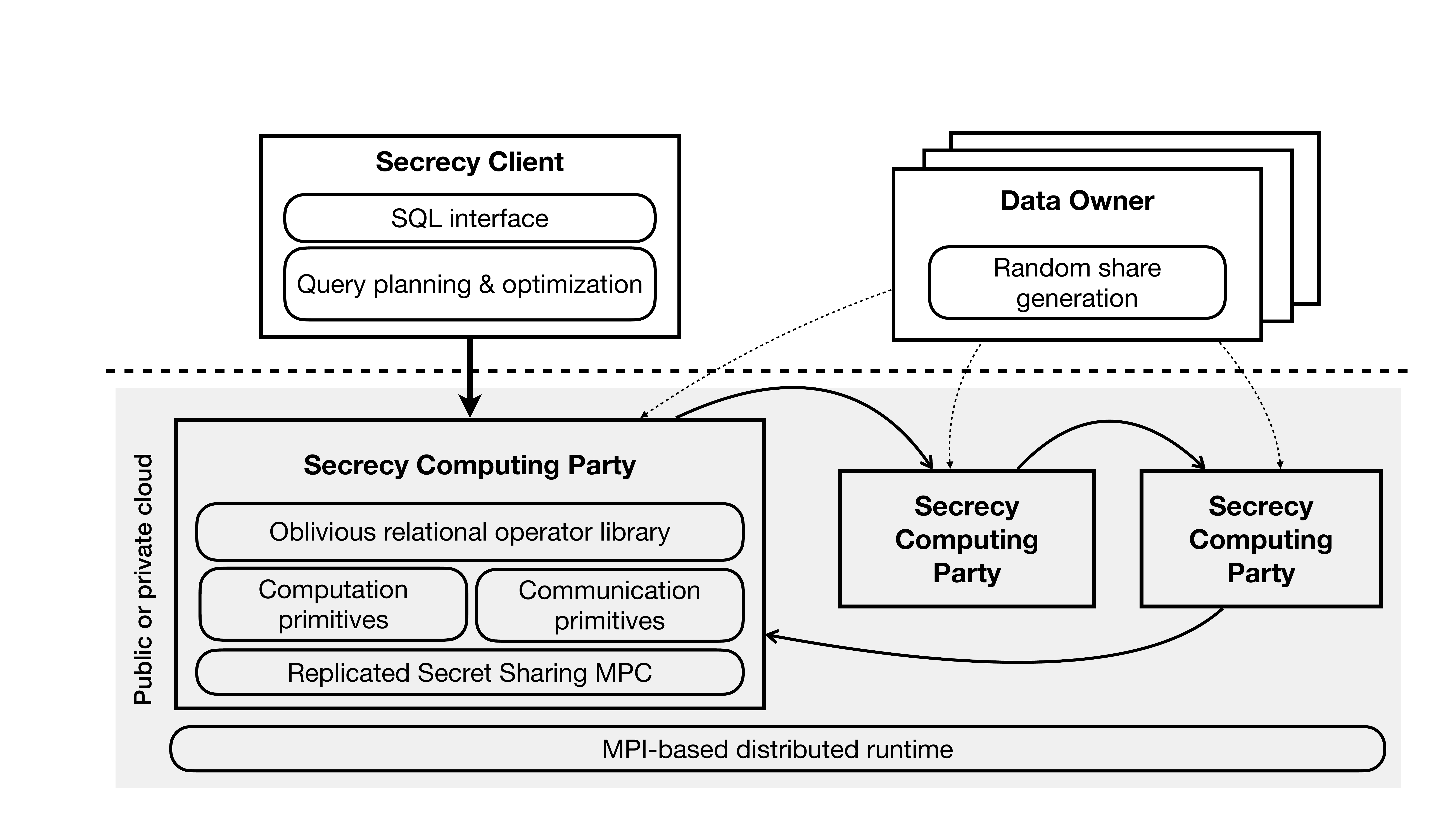}
        \end{minipage} 
        \vspace{-3mm}
        \caption{The \ours system consists of (i) a client application that can be used by data analysts to submit SQL queries and (ii) three computing parties that execute queries under MPC.}\label{fig:stack}
    \end{figure}

Despite a rich open-source ecosystem of general-purpose MPC frameworks~\cite{DBLP:conf/sp/HastingsHNZ19}, we found that existing tools either lack support for general relational operations or cannot effectively amortize network I/O. For these reasons, we implemented \ours in C/C++, entirely from scratch.  We designed our secure primitives to operate directly on relations and we also built a library of general oblivious relational operators that can be combined into arbitrary~query~plans.

\stitle{System overview.}
Figure~\ref{fig:stack} shows the \ours system architecture and software stack.  Data analysts submit queries through a client application that exposes a SQL interface and provides a query planner that performs query rewriting and cost-based optimization (cf. \S~\ref{sec:optimizations}).
Data owners use the secret-sharing generation module to create and distribute random shares of their data to the computing parties.  Computing parties can be deployed on premises, in a federated cloud, or across multiple independent clouds.  Their software stack consists of (i) a custom implementation of the replicated secret sharing protocol, (ii) a library of secure computation and communication primitives, and (iii) a library of oblivious relational operators. The distributed runtime and communication layer are based on MPI~\cite{mpi}. 
Each computing party is a separate MPI process that handles both computation and communication.  

\stitle{Oblivious relational operators.}
\ours supports secure projection, selection, group-by with aggregation, order-by, distinct, and general-purpose theta joins and semi-joins.  Relational operators and secure primitives are designed to process table rows in batches.
The batch size is configurable and allows \ours to compute expensive operators, such as joins, with full control over memory requirements. While batching does not reduce the total number of operations, we leverage it to compute on large inputs without running out of memory or switching to a disk-based evaluation.

\stitle{Query planning and execution.}
Upon startup, the parties establish connections to each other and learn the process IDs of other parties.  
Next, they receive input shares for each base relation from the data owners. 
Queries are specified either in SQL (and go through query planning) or in a low-level API that allows seamless operator composition by abstracting communication and other MPC details.  \ours's Volcano-style planner uses a typical bottom-up dynamic programming algorithm that applies the rules of \S~\ref{sec:optimizations} to generate equivalent plans. It then selects the plan with the lowest total cost (cf. \S~\ref{sec:plan-costs}).  To compute a query, parties execute the same oblivious physical plan on their random data shares and return the results to the designated client.  We use a \texttt{64-bit} data representation for shares, so $\ell = 64$ (cf. \S~\ref{sec:costs}). 
    \begin{figure*}[t]
    \centering
        \begin{minipage}{.48\textwidth}
            \begin{subfigure}{\textwidth}
            \centering
            \subcaption{LAN}\label{baseline-LAN}
  		\includegraphics[width=\linewidth]{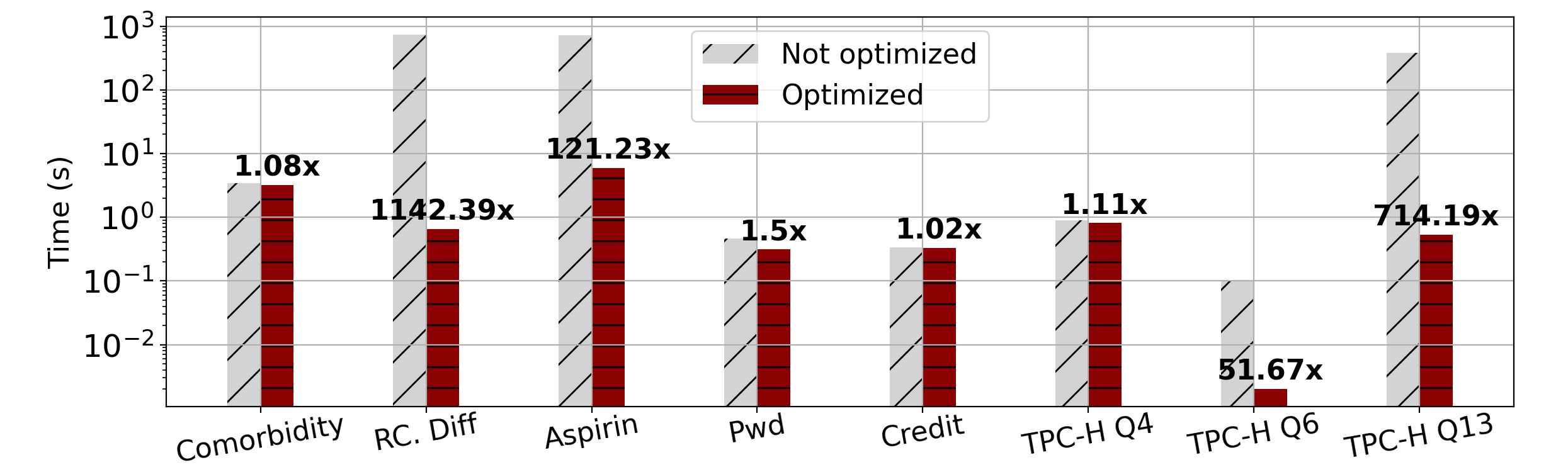}
            \end{subfigure}
        \end{minipage}
        \hfill
         \begin{minipage}{.48\textwidth}
            \begin{subfigure}{\textwidth}
            \centering
            \subcaption{WAN}\label{baseline-WAN}
  		\includegraphics[width=\linewidth]{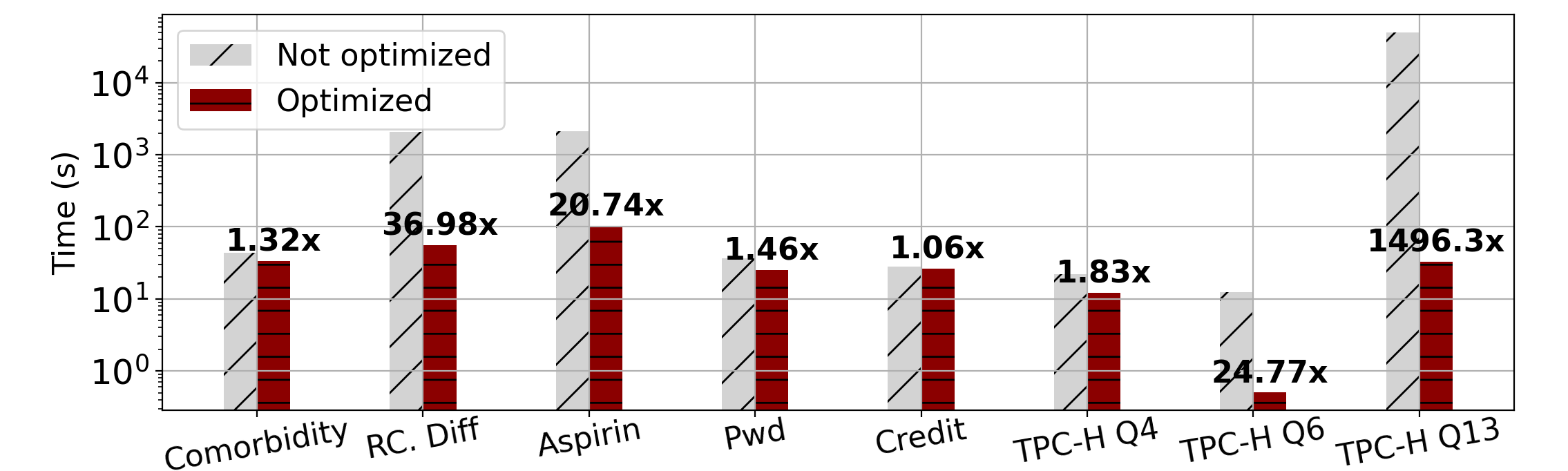}
            \end{subfigure}
        \end{minipage}
         \vspace{-3mm}
         \caption{\small Performance gains of \ours's optimizations over non-optimized plans for real and synthetic queries. Logical and physical optimizations result in over $100\times$ lower execution times, while secret-sharing optimizations improve performance by~up~to~$52\times$.}\label{fig:baseline}
    \end{figure*}

\section{Experimental evaluation}\label{sec:evaluation}
Our experimental evaluation is structured into three parts:

\stitle{Benefits of query optimization.} In \S~\ref{sec:exp-optimizations}, we evaluate the benefits of \ours's optimizations on eight real and synthetic queries. We show that \ours's cost-based optimizer reduces the runtime of complex queries by up to three orders of magnitude both in the LAN and the WAN setting.

\stitle{Performance on real and synthetic queries.} 
In \S~\ref{sec:scalability} we evaluate \ours's performance as input sizes grow.  We use queries that include selections,  group-by,  distinct,  semi-join, and theta-joins with both equality and inequality predicates.  Our results demonstrate that \ours can scale to millions of input rows and evaluate complex queries in reasonable time with modest use of resources.

\stitle{Micro-benchmarks.} In \S~\ref{sec:exp-benefits},  we evaluate individual logical, physical, and secret-sharing optimizations on the three queries of Sections~\ref{sec:logical}-\ref{sec:mpc_optimization}. Our results demonstrate that pushing down blocking operators reduces execution time by up to $1000\times$ and enables queries to scale to $100\times$ larger inputs.  Further, we show that operator fusion and dual sharing  improve execution time by an order of magnitude in the WAN setting.

\stitle{Comparison with state-of-the-art frameworks.} In \S~\ref{sec:eval_comparison}, we compare \ours with
SMCQL~\cite{Bater2017SMCQL} and the 2-party semi-honest version of EMP~\cite{emp-toolkit} with $N=2$ and $T=1$ (cf. \S~\ref{sec:guarantees}). 
We show that \ours outperforms them both and can comfortably process much larger datasets within the same amount of time.

\stitle{Performance of relational operators.} In \S~\ref{sec:exp-relational}, we present performance results for individual relational operators.
 We show that \ours's batched operator implementations are efficient and that by properly adjusting the batch size, they can comfortably scale to millions of input rows without running out of memory.

\stitle{Performance of \ours's primitives.} Finally, in Section~\ref{sec:exp-micro}, we drill down and evaluate individual secure computation and communication primitives that relational operators rely upon. We empirically verify the theoretical cost analysis of \S~\ref{sec:cost_primitives}, evaluate the scalability of primitives, and quantify the positive effect that message batching has on the performance of I/O-heavy operations.

\subsection{Evaluation setup}

We use two cloud deployments: (i) \texttt{AWS-LAN} uses an \textit{EC2 r5.xlarge} instance per party in the \texttt{us-east-2} region,  and (ii) \texttt{AWS-WAN} distributes parties across \texttt{us-east-2} (Ohio), \texttt{us-east-1} (Virginia), and \texttt{us-west-1} (California). VMs have 32GB of memory and run Ubuntu \texttt{20.04}, \texttt{C99}, \texttt{gcc 5.4.0}, and \texttt{MPICH 3.3.2}. 
We designate one party as the data owner that distributes shares and reveals results. 
Measurements are averaged over at least three runs and plotted in log-scale, unless otherwise specified.

\stitle{Queries.} We use 11 queries for evaluation, including five real-world queries from previous MPC works~\cite{Bater2017SMCQL, Volgushev2019Conclave, bater2018shrinkwrap, Poddar2021Senate, bater2020saqe}. 
Three are medical queries~\cite{Bater2017SMCQL}: \emph{Comorbidity} returns the ten most common diagnoses of individuals in a cohort,  \emph{Recurrent C.Diff.} returns the distinct ids of patients who have been diagnosed with \texttt{cdiff} and have two consecutive infections between 15 and 56 days apart, and \emph{Aspirin Count} returns the number of patients who have been diagnosed with heart disease and have been prescribed aspirin after the diagnosis was made.
We also use queries from other MPC application areas~\cite{Poddar2021Senate}: \emph{Password Reuse} asks for users with the same password across different websites, while \emph{Credit Score} asks for persons whose credit scores across different agencies have significant discrepancies in a particular year. In addition to the real-world queries, we use the TPC-H queries (\textbf{Q4}, \textbf{Q6}, \textbf{Q13})~\cite{tpch} that have been used in SAQE~\cite{bater2020saqe}. 
Finally, to evaluate the performance gains from each optimization in isolation, we use \textbf{Q1}, \textbf{Q2}, \textbf{Q3} of \S~\ref{sec:logical}-\ref{sec:mpc_optimization}.

\stitle{Datasets.} In all experiments, we use randomly generated tables with 64-bit values. Note that \ours's MPC protocol assumes a fixed-size representation of shares that is implementation-specific and could be increased to any $2^{k}$ value. We also highlight that using randomly generated inputs is no different than using real data, as all operators are oblivious and the data distribution does not affect the amount of computation or communication. No matter whether the input values are real or random, parties compute on secret shares, which are by definition random.    

\subsection{Benefits of query optimization}\label{sec:exp-optimizations}

We compare the performance of eight queries optimized by \ours with that of plans without the optimizations of Section~\ref{sec:optimizations}.  For a fair comparison,  we implement baseline plans using \ours's batched operators. Although this favors the baseline, the communication cost of MPC is otherwise prohibitive and queries cannot scale beyond a few hundred input rows. 
We execute each query plan with $1K$ rows per input relation. For Q4 (resp. Q13), we use $1K$ rows for \texttt{LINEITEM} (resp. \texttt{ORDERS}) and maintain the size ratio with the other input relation as specified in the TPC-H benchmark.  For \emph{Comorbidity}, we use a cohort of $256$ patients. 
We run this experiment on \texttt{AWS-LAN} and \texttt{AWS-WAN} and present the results in Figure~\ref{fig:baseline}. 

In the LAN setting, \ours achieves the highest speedups for \emph{Recurrent C.Diff.}, \emph{Aspirin Count}, and Q13, that is, $1142\times$, $121\times$, and $714\times$ lower execution times, respectively. Optimized plans for these queries leverage join push-up (\emph{Aspirin Count}), fusion (\emph{Recurrent C.Diff.}), and join-aggregation decomposition (Q13). The optimized plans for \emph{Comorbidity}, \emph{Password Reuse}, Q4, and Q6 leverage dual and proactive sharing, achieving up to $52\times$ speedup compared to the baseline. Finally, the \emph{Credit Score} query leverages dual sharing which, in this case, provides a modest improvement.
\ours achieves significant speedups in the wide area, too.  The performance improvement is higher for \emph{Comorbidity}, Q4, and Q13 in the WAN setting, as these queries leverage optimizations that primarily reduce the synchronization cost. We evaluate the benefit of individual optimizations~in~\S~\ref{sec:exp-benefits}.

 \begin{figure*}[t]
    \vspace{-4mm}
    \centering
      \begin{minipage}{.3\textwidth}
            \begin{subfigure}{\textwidth}
            \centering
            \caption{Category A}\label{fig:cat1-queries}
  		\includegraphics[width=\linewidth]{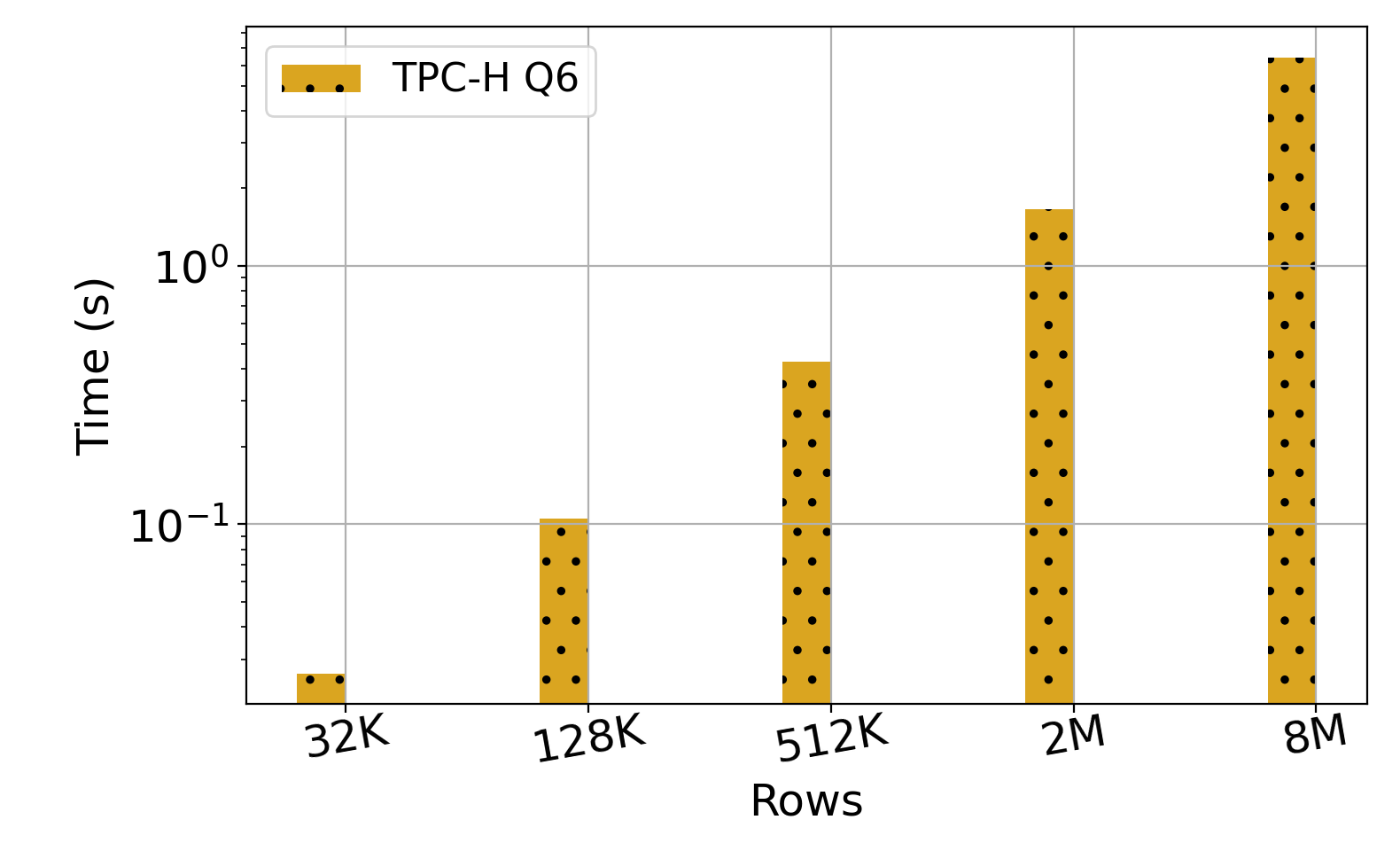}
            \end{subfigure}
        \end{minipage}
        \hfill
        \begin{minipage}{.31\textwidth}
            \begin{subfigure}{\textwidth}
            \centering
            \caption{Category B}\label{fig:cat2-queries}
  		\includegraphics[width=\linewidth]{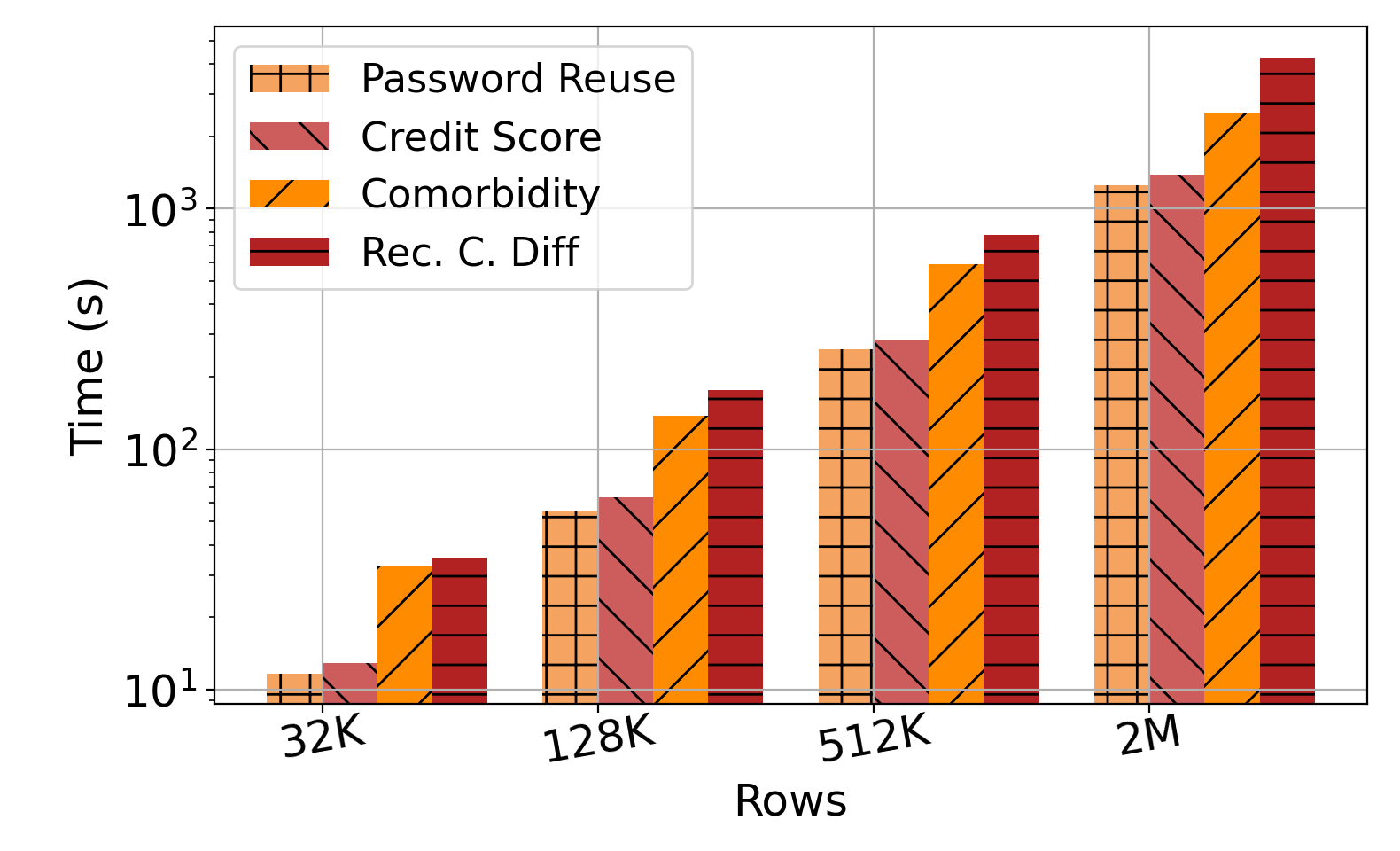}			
            \end{subfigure}
        \end{minipage}
        \hfill
        \begin{minipage}{.3\textwidth}
            \begin{subfigure}{\textwidth}
            \centering
            \caption{Category C}\label{fig:cat3-queries}
  			\includegraphics[width=\linewidth]{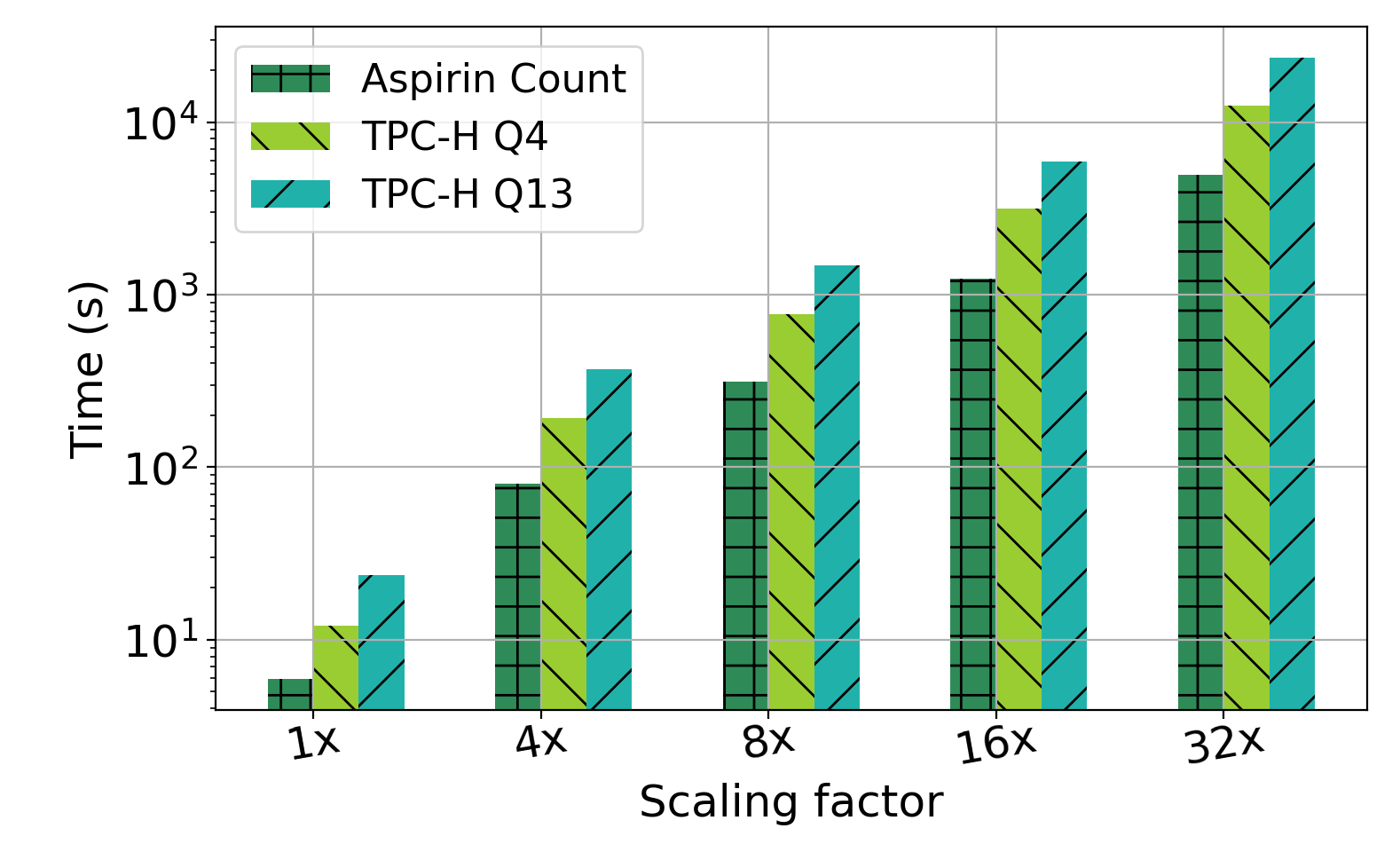}
            \end{subfigure}
        \end{minipage}
         \vspace{-4mm}
         \caption{Scaling behavior of optimized real and synthetic queries on \ours (LAN)}
         \label{fig:scaling}
    \end{figure*}  
    
        \begin{figure*}[t]
    \vspace{-4mm}
    \centering
      \begin{minipage}{.3\textwidth}
            \begin{subfigure}{\textwidth}
            \centering
            \caption{Category A}\label{fig:cat1-queries}
  		\includegraphics[width=\linewidth]{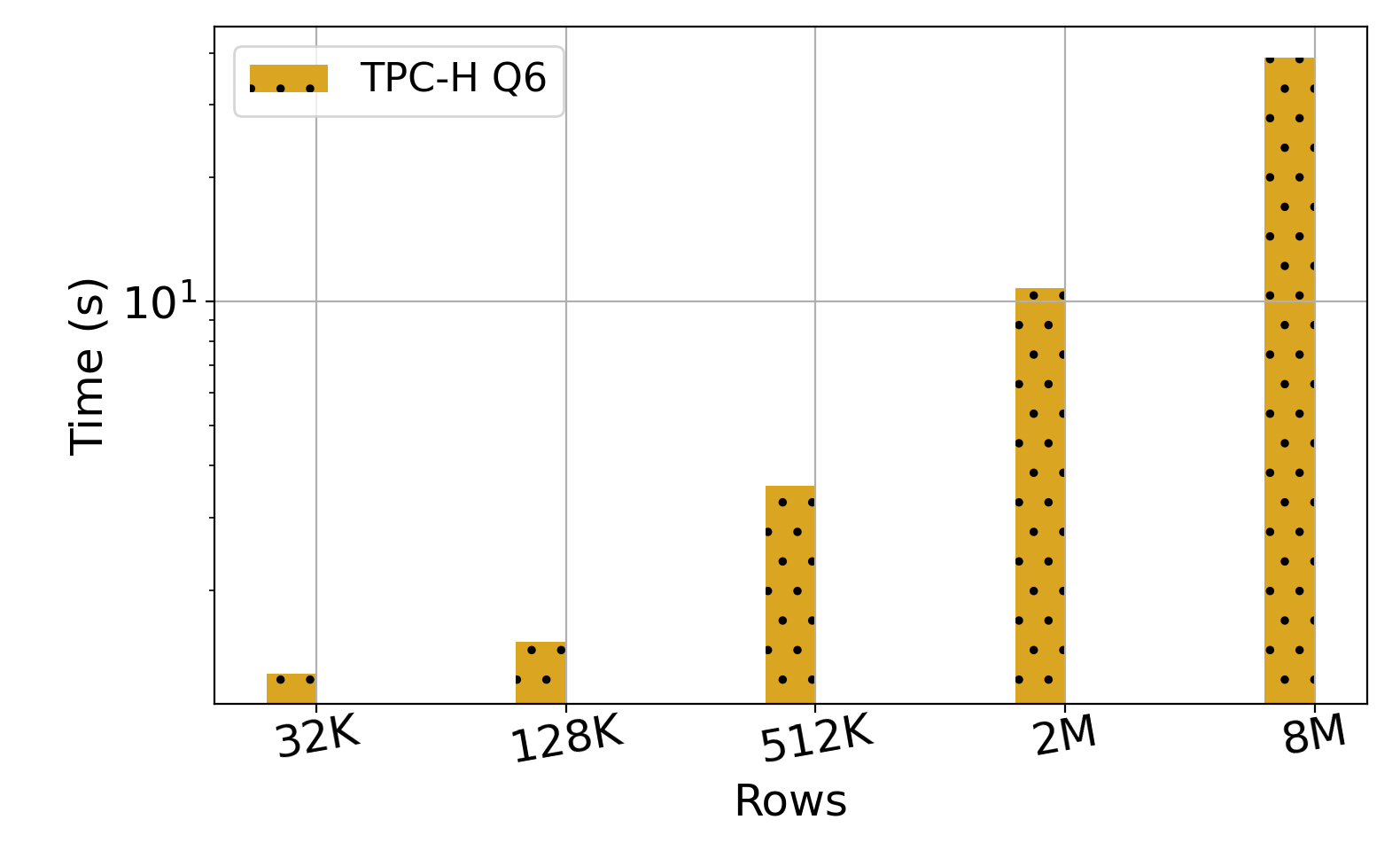}
            \end{subfigure}
        \end{minipage}
        \hfill
        \begin{minipage}{.31\textwidth}
            \begin{subfigure}{\textwidth}
            \centering
            \caption{Category B}\label{fig:cat2-queries}
  		\includegraphics[width=\linewidth]{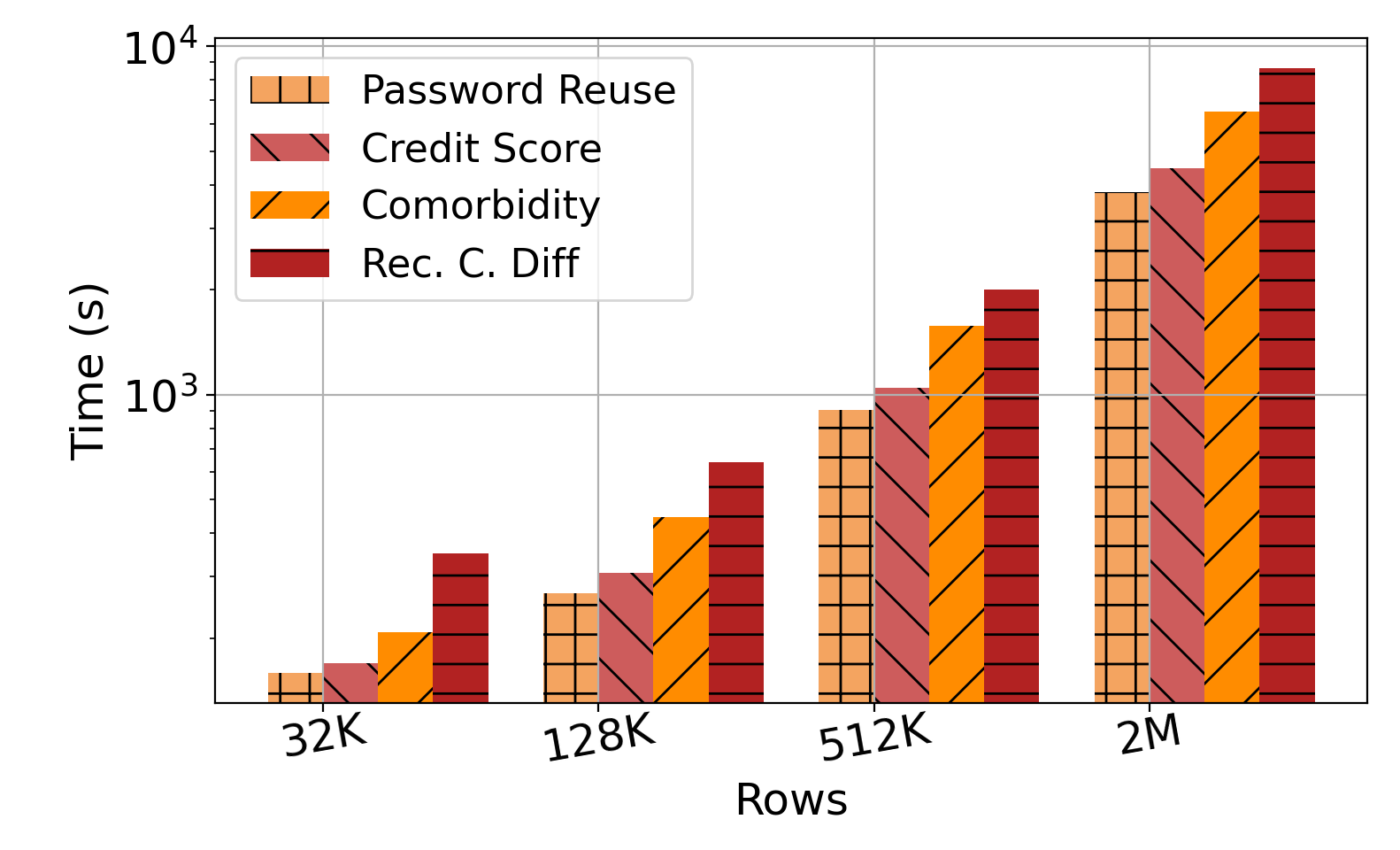}			
            \end{subfigure}
        \end{minipage}
        \hfill
        \begin{minipage}{.3\textwidth}
            \begin{subfigure}{\textwidth}
            \centering
            \caption{Category C}\label{fig:cat3-queries}
  			\includegraphics[width=\linewidth]{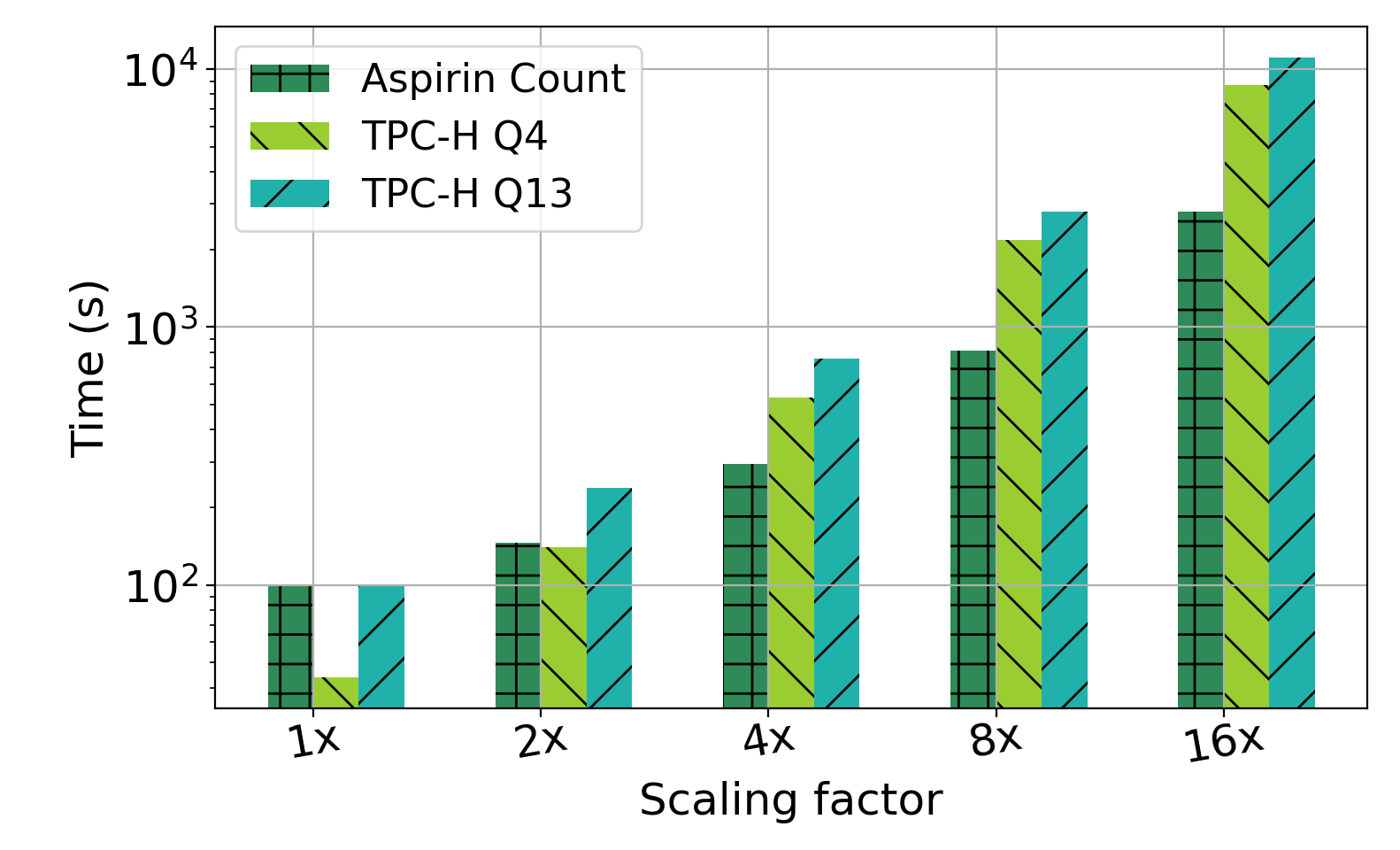}
            \end{subfigure}
        \end{minipage}
         \vspace{-4mm}
         \caption{Scaling behavior of optimized real and synthetic queries on \ours (WAN)}
         \label{fig:scaling-wan}
    \end{figure*}  
    
\subsection{Performance on real and synthetic queries}\label{sec:scalability}
We now run the optimized plans with increasing input sizes and report total execution time in \texttt{AWS-LAN} and \texttt{AWS-WAN}. 
For these experiments, we group queries into three categories of increasing complexity. \emph{Category~A} includes queries with selections and global aggregations, 
\emph{Category~B} includes queries with select and group-by or distinct operators,
and \emph{Category~C} includes queries with select, group-by and (semi-)join operators.
Figures~\ref{fig:scaling} and \ref{fig:scaling-wan} present the results in LAN and WAN respectively. 

Q6 in \emph{Category A} consists of five selections and a global aggregation. It requires minimal inter-party communication that is independent of the input relation cardinality.
As a result, it scales comfortably to large inputs and takes $\sim6s$ (resp. $\sim39s$) for $8M$ rows in LAN (resp. WAN). 

Queries in \emph{Category B} scale to millions of input rows as well. 
The cost of these queries is dominated by the oblivious group-by and distinct operators.  
At $2M$ rows, \emph{Recurrent C.Diff.} completes in $\sim1.2h$ in LAN and $\sim2.4h$ in WAN. For the same input, \emph{Password Reuse} completes in $\sim20min$ in LAN and $\sim1h$ in WAN.

The cost of queries in \emph{Category C} is dominated by joins and semi-joins. 
The size ratio between the two inputs of each query is different: 
for Q4 and Q13, we use the ratio specified in the TPC-H benchmark whereas, for \emph{Aspirin Count}, we use inputs of equal size. 
In Figure~\ref{fig:cat3-queries}, Scaling factor $1\times$ corresponds to $1K$ rows for the small input. As we increase the input sizes, we always keep their ratio fixed.
At scaling factor $32\times$, the most expensive query is Q13, which is optimized with join-aggregation decomposition and takes $\sim6.5h$ on $295K$ rows. At the same scaling factor, Q4 completes in $\sim3.4h$ on $164K$ rows, and \emph{Aspirin Count} in $\sim1.3h$.
For WAN, at scaling factor $16\times$, Q13 takes $\sim3h$. At the same scaling factor, Q4 completes in $\sim2.4h$ and \emph{Aspirin Count} in $\sim46min$.

While MPC protocols remain highly expensive for real-time queries, our results demonstrate that offline collaborative analytics on medium-sized datasets entirely under MPC are viable.\vspace{-2mm} 

\begin{figure*}[t]
    \centering
            \begin{minipage}{.23\textwidth}
            \begin{subfigure}{\textwidth}
            \centering
            \caption{Distinct-join reordering (LAN)}\label{fig:distinct-join-opt}
            \includegraphics[width=\textwidth]{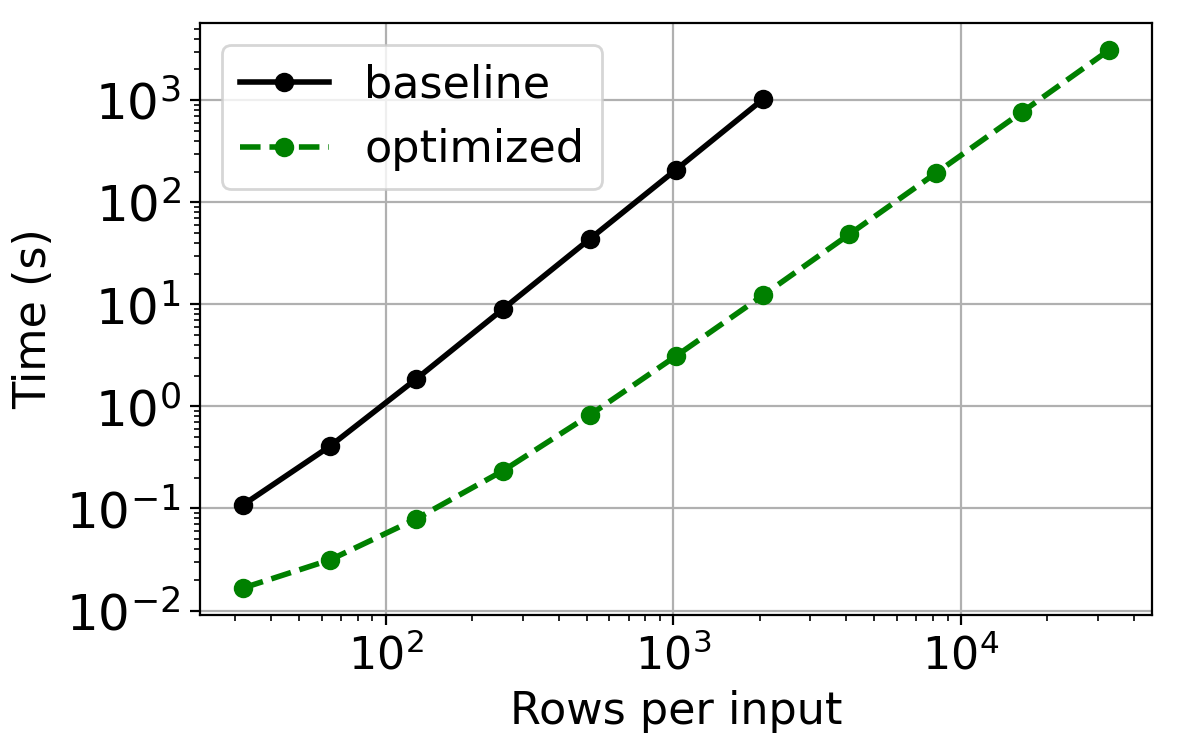}
            \end{subfigure}
        \end{minipage}
        \hfill
        \begin{minipage}{.23\textwidth}
            \begin{subfigure}{\textwidth}
             \centering
             \caption{Join-Aggr. decomposition (LAN)}\label{fig:group-by-opt}
            \includegraphics[width=\textwidth]{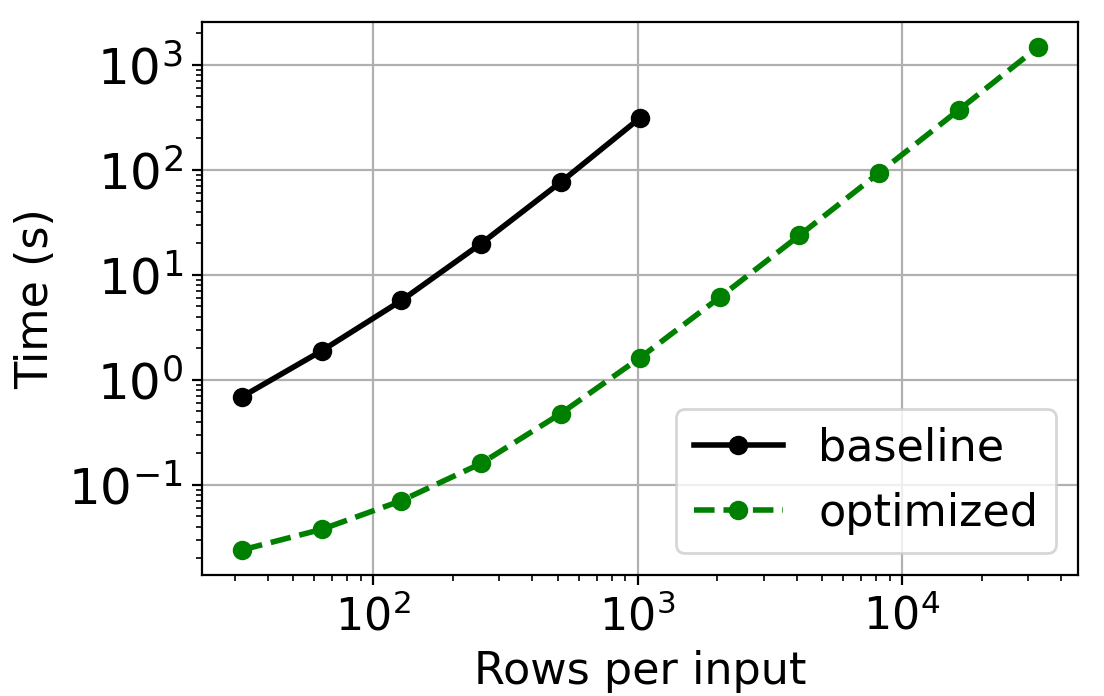}
            \end{subfigure}
        \end{minipage}
        \hfill
        \begin{minipage}{.23\textwidth}
            \begin{subfigure}{\textwidth}
            \centering
            \caption{Select-Distinct fusion (WAN)}\label{fig:select-distinct-opt}
            \includegraphics[width=\textwidth]{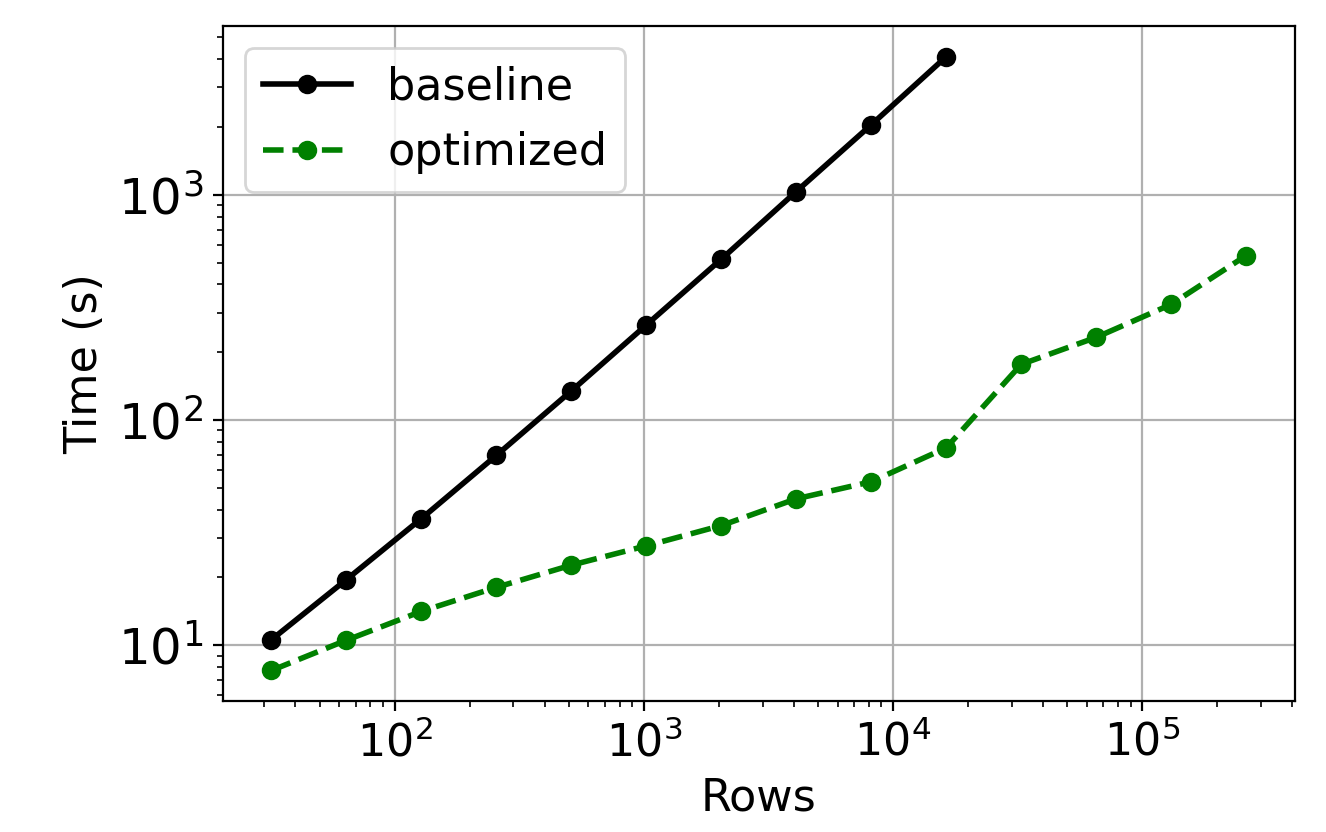}
            \end{subfigure}
        \end{minipage}
          \hfill
        \begin{minipage}{.23\textwidth}
            \begin{subfigure}{\textwidth}
            \centering
            \caption{Dual sharing in Group-by (WAN)}\label{fig:dual-group-by-opt}
            \includegraphics[width=\textwidth]{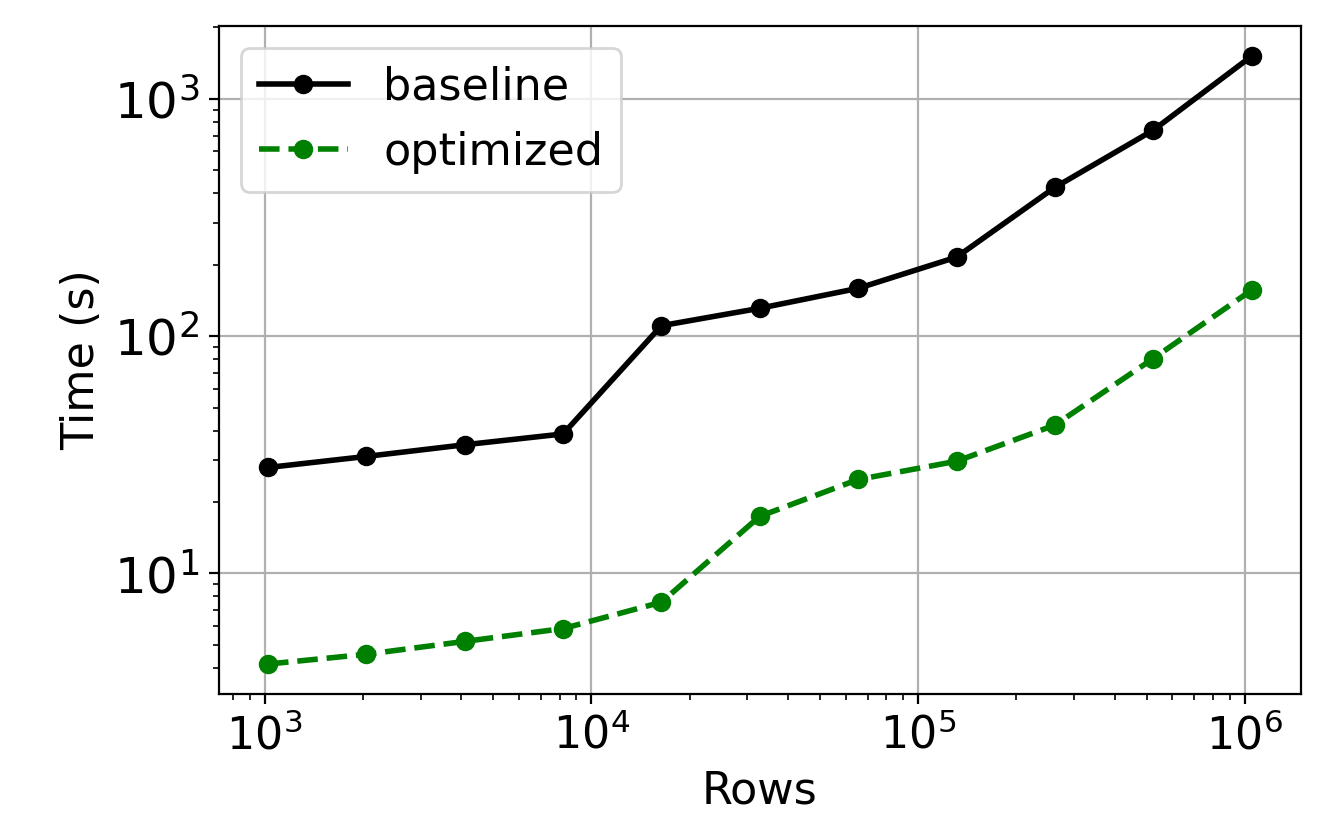}
            \end{subfigure}
        \end{minipage}
         \vspace{-3mm}
         \caption{Performance improvement of individual optimizations applied by the \ours planner.}\label{fig:benefits}
\end{figure*}

\subsection{Micro-benchmarks}\label{sec:exp-benefits}
We now use the queries of \S~\ref{sec:optimizations} (Q1, Q2, Q3) to evaluate the impact of \ours's optimizations in isolation. We run each query with and without the particular optimization and measure total execution time. 
Distinct-join reordering and join-aggregation decomposition primarily reduce the operation cost and we evaluate them in \texttt{AWS-LAN}.  Fusion and dual sharing reduce the synchronization cost and we evaluate them in \texttt{AWS-WAN}.
Figure~\ref{fig:benefits} shows~the~results. 

\stitle{Distinct-Join reordering.}
The optimized plan of Q1 pushes the \texttt{JOIN} after \texttt{DISTINCT} and, thus, only sorts a relation of $n$ rows instead of $n^{2}$. Figure~\ref{fig:distinct-join-opt} shows that the optimized plan is up to two orders of magnitude faster than the baseline, which runs out of memory for even modest input sizes.

\stitle{Join-Aggregation decomposition.}
The baseline plan of Q2 materializes the result of the join and then applies the grouping and aggregation.
Instead, the optimized plan decomposes the aggregation in two phases (cf.~\S~\ref{sec:decomposition}). 
As shown in Figure~\ref{fig:group-by-opt}, this optimization provides up to two orders of magnitude lower execution time than that of the baseline plan. Further, the baseline plan runs out of memory for inputs larger than $1K$ rows. 

\stitle{Operator fusion.}
The baseline plan of Q3 applies the oblivious selection before \texttt{DISTINCT}, while the optimized plan fuses the two operators and performs the \texttt{DISTINCT} computation in bulk (cf.~\S~\ref{sec:distinct-fusion}). Figure~\ref{fig:select-distinct-opt} shows that this optimization provides up to $44\times$ speedup for large inputs and allows the query to scale to much larger inputs.

\stitle{Dual sharing.}
We also evaluate \ours's ability to switch between arithmetic and boolean sharing to reduce communication costs for certain operations. For this experiment, we compare the run-time of the optimized \texttt{GROUP-BY-COUNT} operator (cf.~\S~\ref{sec:mpc_optimization}) to that of a baseline operator that uses boolean sharing only and, hence, relies on the ripple-carry adder to compute the \texttt{COUNT}. Figure~\ref{fig:dual-group-by-opt} plots the results. The baseline operator is $10\times$ slower than the optimized one, as it requires $64$ additional rounds of communication per~input~row. 

\subsection{Comparison with other MPC frameworks}\label{sec:eval_comparison}
Existing 3-party frameworks \cite{DBLP:conf/sp/HastingsHNZ19} are either proprietary, e.g. \cite{DBLP:conf/esorics/BogdanovLW08}, or they only support specific operators, such as unique-key joins~\cite{aby3_github,DBLP:conf/ccs/MohasselRR20}, that cannot be used for any of the queries we consider. 
For the comparisons of this section, we choose SMCQL (the ORAM-based version) as the only open-source relational framework with semi-honest security and no information leakage. We also choose EMP since it is used by all recent systems, namely Shrinkwrap~\cite{bater2018shrinkwrap}, SAQE~\cite{bater2020saqe}, a new version of SMCQL, and Senate~\cite{Poddar2021Senate}. Although none of these systems is publicly available, they all build their relational MPC engines on top of EMP.
We stress that EMP and SMCQL use 2-party protocols whose threat models are not directly comparable with \ours's.
The purpose of these experiments is to showcase the end-to-end performance of the available solutions to relational MPC and not to compare the underlying protocols.

\begin{table}[t]
\scriptsize
\centering
\begin{tabular}{c|r|r|r|}
\cline{2-4}
\multicolumn{1}{l|}{} &
  \multicolumn{1}{c|}{\textbf{Comorbidity}} &
  \multicolumn{1}{c|}{\textbf{Recurrent C. Diff.}} &
  \multicolumn{1}{c|}{\textbf{Aspirin Count}} \\ \hline
  \hline
\multicolumn{1}{|c|}{\textbf{SMCQL}} &
$197s$& $804s$ & $796s$\\ \hline
\multicolumn{1}{|c|}{\textbf{\ours}} &
$0.083s$& $0.092s$ & $0.171s$\\ \hline
\bottomrule
\end{tabular}
\caption{SMCQL and \ours execution times for the three medical queries from~\cite{Bater2017SMCQL} on 25 tuples per input relation.}
\label{tab:smcql-results}\vspace{-3mm}
\end{table}
        
\begin{figure}[t]
    \centering
        \begin{minipage}{.24\textwidth}
            \begin{subfigure}{\textwidth}
            \centering
  		\includegraphics[width=\linewidth]{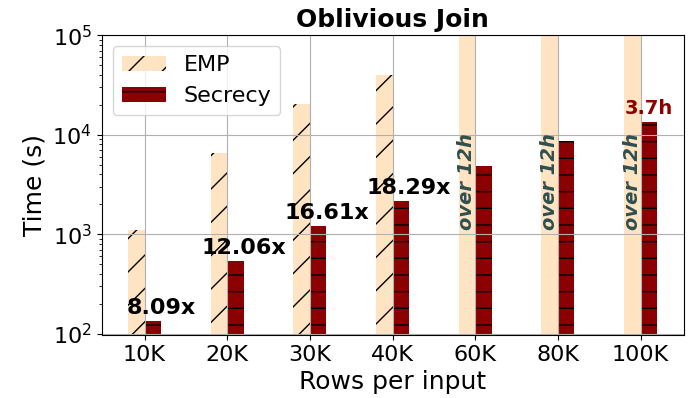}
            \end{subfigure}
        \end{minipage}
        \hfill
        \begin{minipage}{.23\textwidth}
            \begin{subfigure}{\textwidth}
            \centering
  			\includegraphics[width=\linewidth]{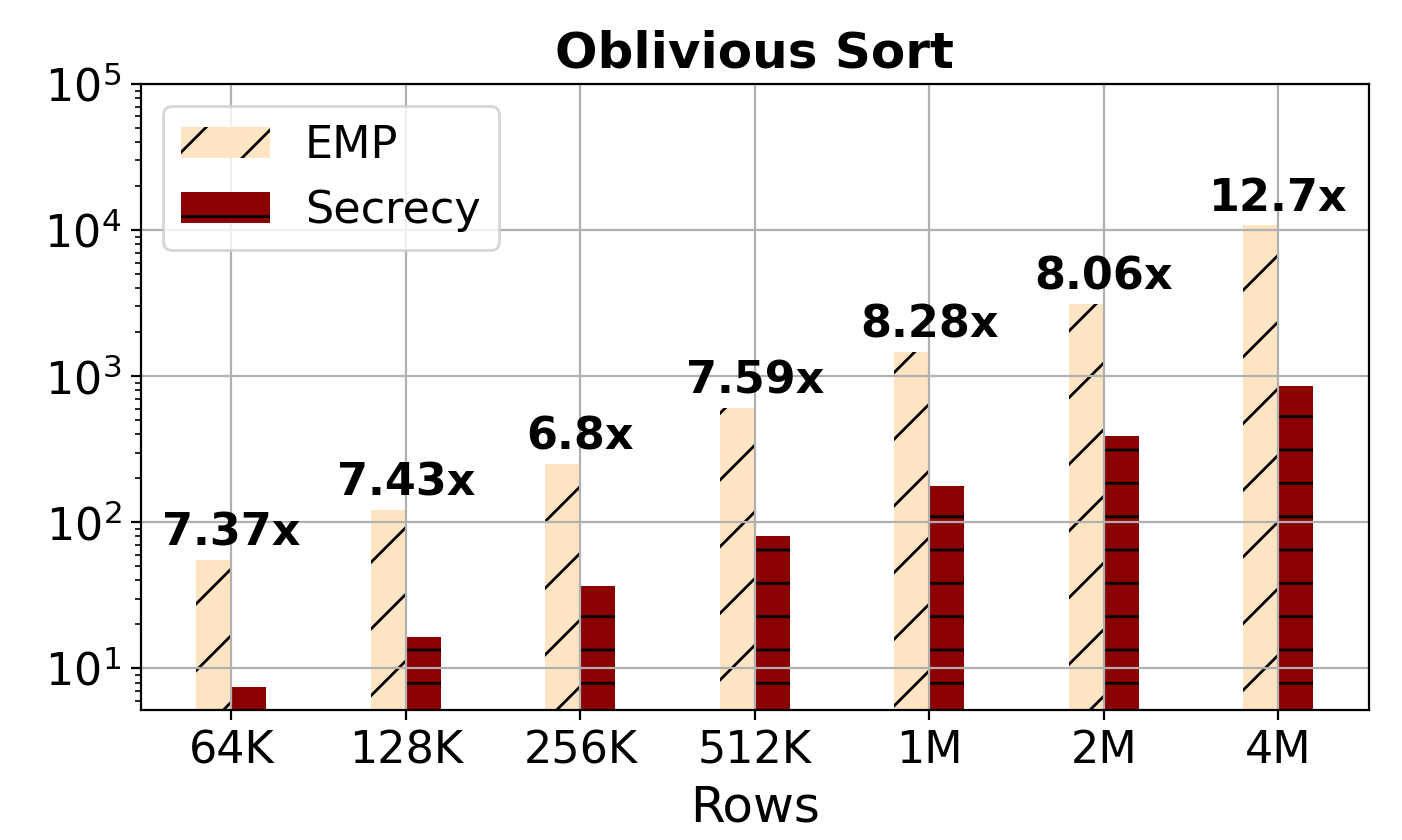}
            \vspace{-5mm}
            \end{subfigure}
        \end{minipage}
         \vspace{-3mm}
         \caption{Performance comparison between EMP and \ours (LAN)}
        	\label{fig:emp-results}
        	\vspace{-3mm}
\end{figure}

\begin{figure}[t]
    \centering
        \begin{minipage}{.24\textwidth}
            \begin{subfigure}{\textwidth}
            \centering
  		\includegraphics[width=\linewidth]{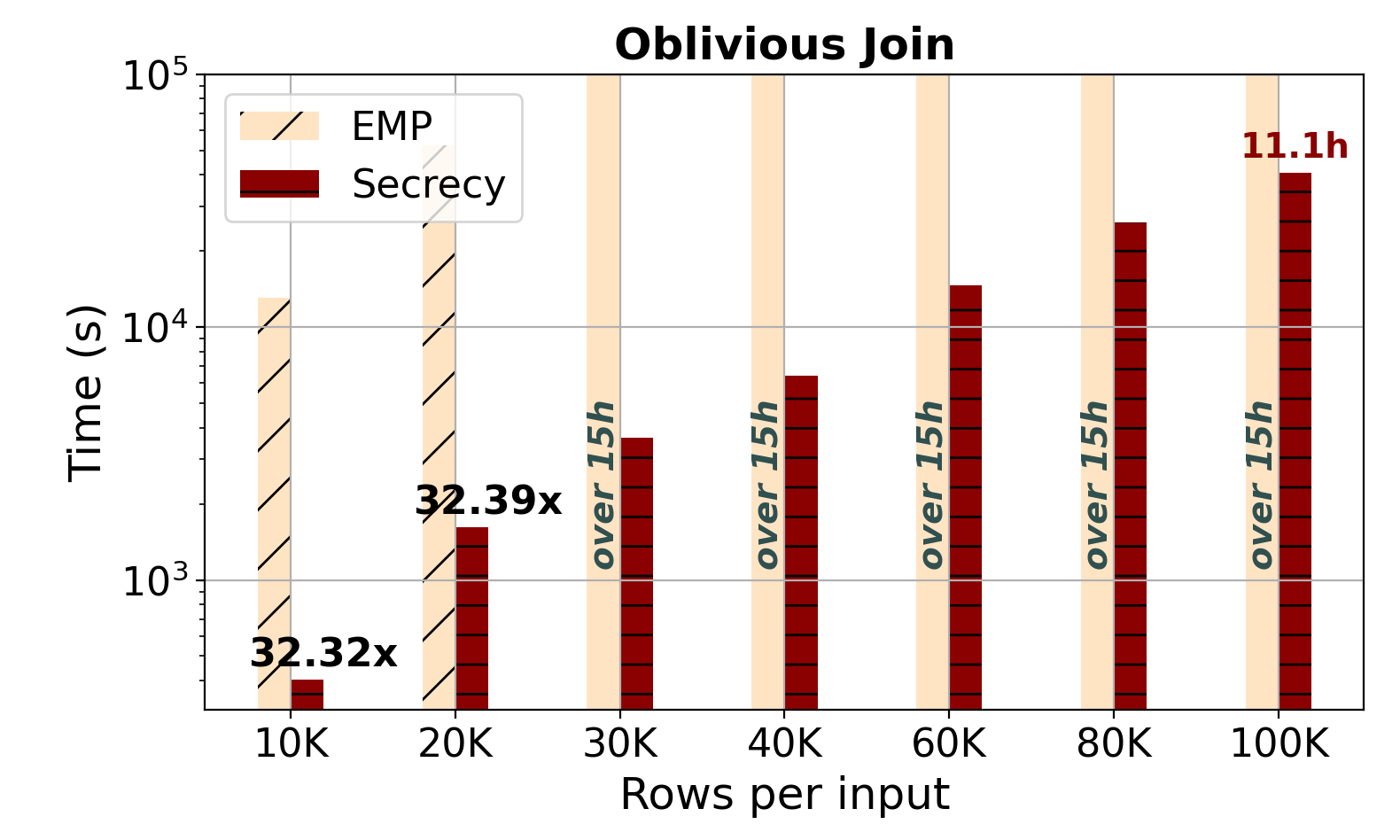}
            \end{subfigure}
        \end{minipage}
        \hfill
        \begin{minipage}{.23\textwidth}
            \begin{subfigure}{\textwidth}
            \centering
  			\includegraphics[width=\linewidth]{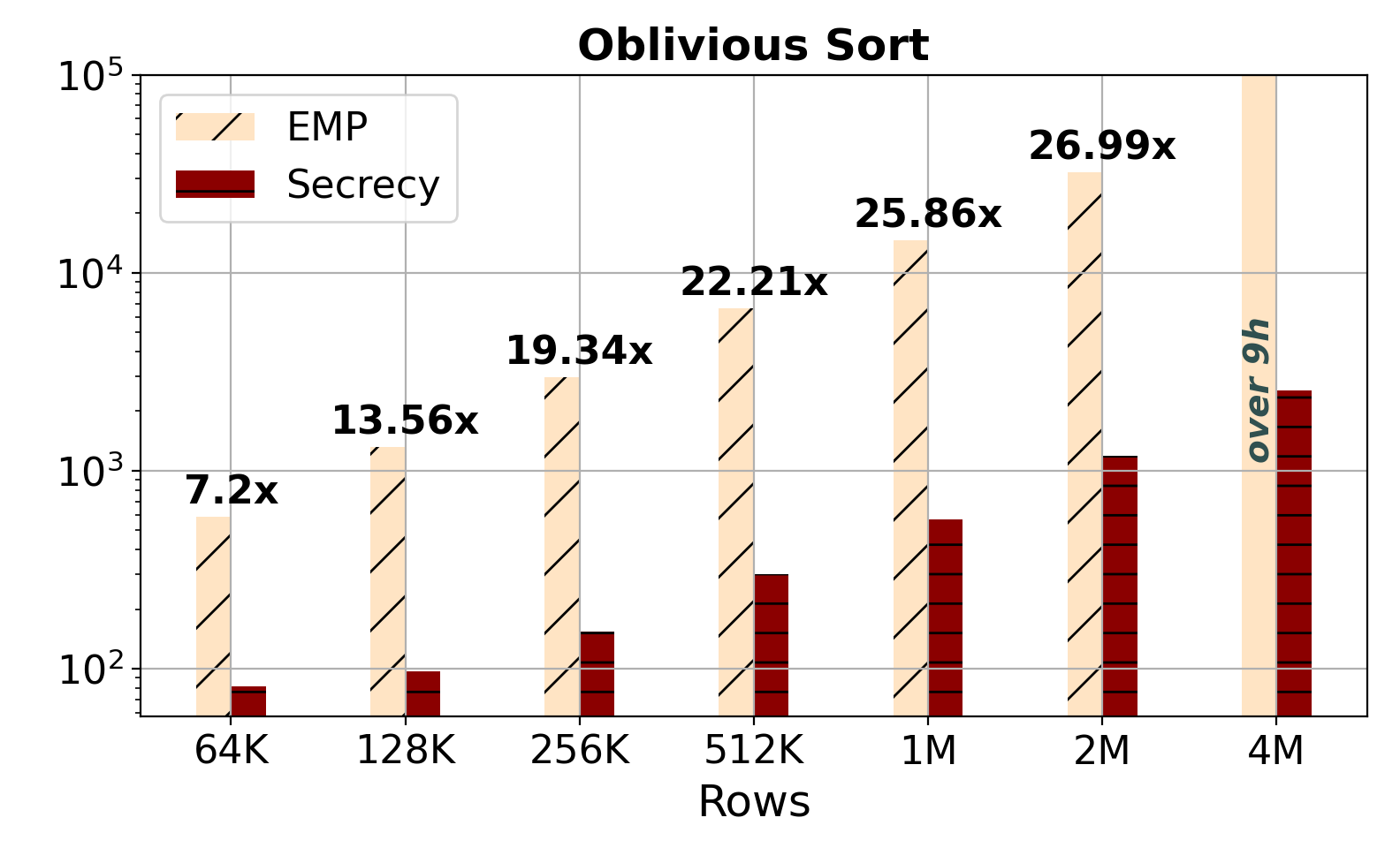}
            \vspace{-5mm}
            \end{subfigure}
        \end{minipage}
         \vspace{-3mm}
         \caption{Performance comparison between EMP and \ours (WAN)}
        	\label{fig:emp-results-wan}
        	\vspace{-3mm}
\end{figure} 

\stitle{Comparison with SMCQL.} In the first set of experiments, we aim to reproduce the results presented in SMCQL~\cite[Fig.~7]{Bater2017SMCQL} in our experimental setup. We run the three medical queries on SMCQL and \ours, using a sample of $25$ rows per data owner (50 in total), and present the results in Table~\ref{tab:smcql-results}. We use the plans and default configuration of protected and public attributes, as in the SMCQL project repository. 
\ours is over $2000\times$ faster than SMCQL in all queries, even though SMCQL pushes operators outside the MPC boundary by allowing parties (that are also data owners) to execute part of the computation on their plaintext data. 

   \begin{figure}[t]
    \centering
        \begin{minipage}{.21\textwidth}
            \begin{subfigure}{\textwidth}
            \centering
            \includegraphics[width=\textwidth]{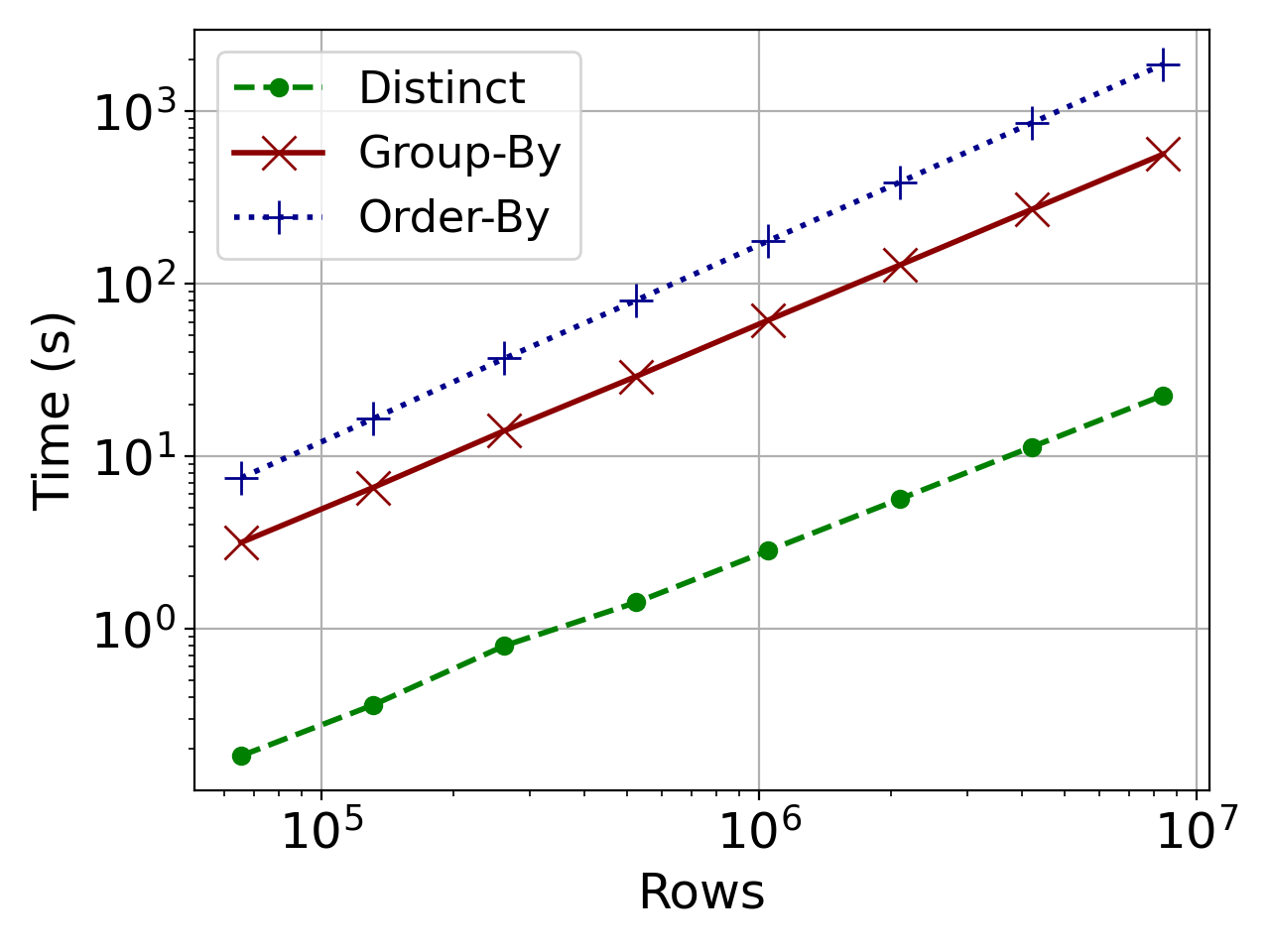}
            \vspace{-6mm}
            \caption{Unary operators}\label{fig:group-distinct-order}
            \end{subfigure}
        \end{minipage}
        \hspace{5mm}
        \begin{minipage}{.21\textwidth}
            \begin{subfigure}{\textwidth}
            \centering
            \includegraphics[width=\textwidth]{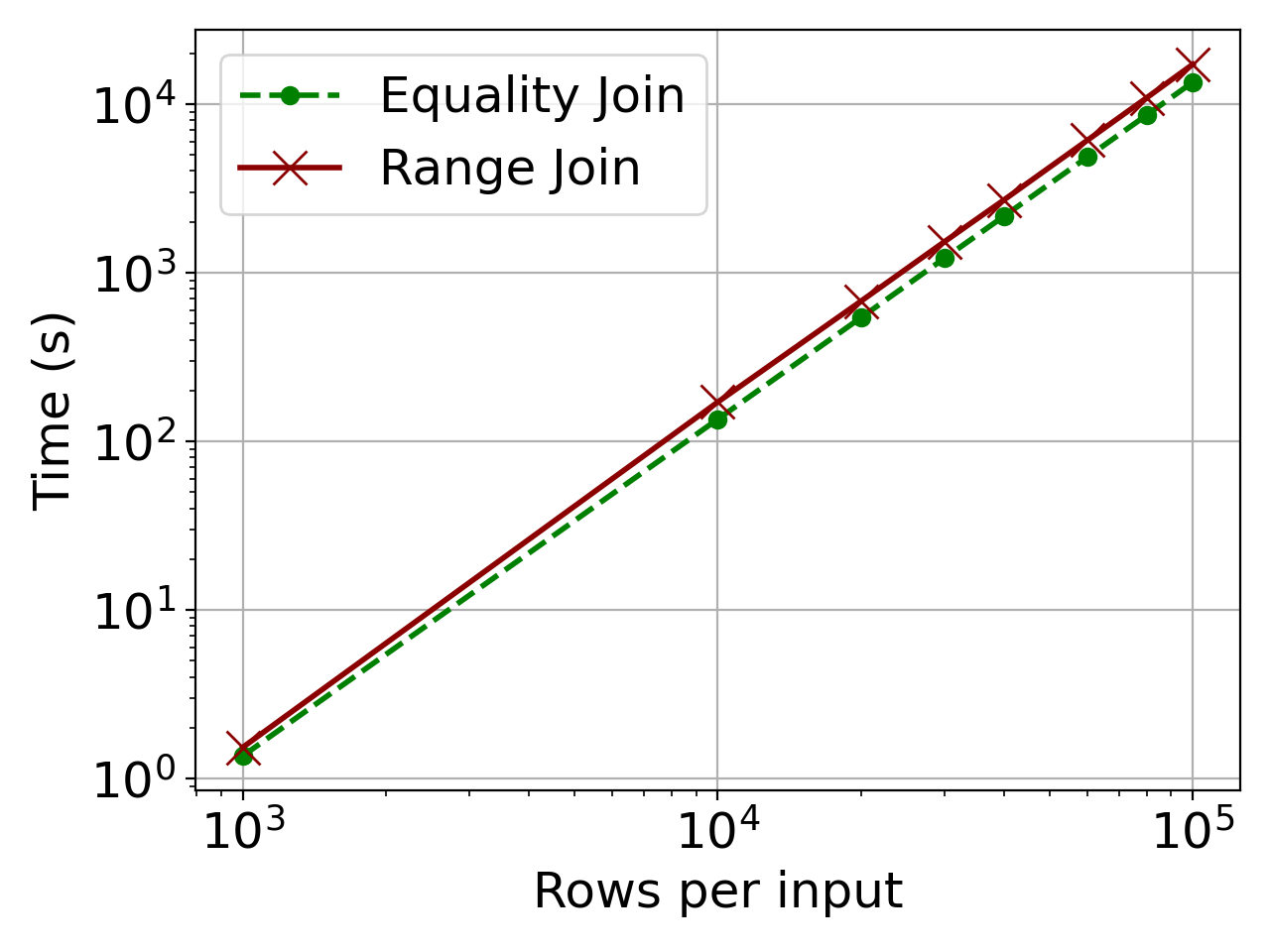}
            \vspace{-6mm}
            \caption{Join operators}\label{fig:joins}
            \end{subfigure}
        \end{minipage}
          \hspace{5mm}
         \caption{Performance of oblivious relational operators in \ours}
         \label{fig:micro-relational} \vspace{-2mm}
    \end{figure}

\stitle{Comparison with EMP.}
EMP is a general-purpose MPC framework and does not provide implementations of relational operators or query planning. 
For these experiments, we use two individual operators: an oblivious join operator based on the sample program from the SoK project~\cite{sok-project} and an oblivious sort from the EMP repository~\cite{emp-toolkit}.  Both these operators have the same asymptotic complexity with the respective \ours operators. 
Figure~\ref{fig:emp-results} show the results in \texttt{AWS-LAN}.
For joins, we use inputs of the same cardinality and increase the size from $10K$ to $100K$ rows per input. We cap the time of these experiments to $12h$. Within the experiment duration, EMP can evaluate joins on up to $40K$ rows per input (in $\sim11h$). \ours is $18\times$ faster for the same input size and can process up to $100K$ rows per input in less than $4h$. 
The performance gap between \ours and EMP on sort is also significant. In this case, \ours is up to $12.7\times$ faster ($\sim 3h$ vs $14min$ for $4M$ input~rows). 

In \texttt{AWS-WAN}, the performance difference between \ours and EMP is even more profound due to \ours's effective message batching. Figure~\ref{fig:emp-results-wan} shows the results. Within the experiment duration ($15h$), EMP can evaluate joins on up to $20K$ rows per input (in $\sim14.5h$). \ours is $32\times$ faster for the same input and can process up to $100K$ rows per input in $\sim11h$. For sort, the performance difference between the two systems is similar: \ours is up to $27\times$ faster ($\sim8.7h$ vs $\sim33min$ for $2M$ rows) and can process $4M$ rows~in~$\sim42min$.
 
\subsection{Performance of relational operators}\label{sec:exp-relational}
The next set of experiments evaluates the performance of oblivious relational operators in \ours. We perform \texttt{DISTINCT}, \texttt{GROUP-BY}, \texttt{ORDER-BY}, and \texttt{JOIN} (equality and range) on relations of increasing size and measure the total execution time per operator in \texttt{AWS-LAN}.  We empirically verify the cost analysis of Section~\ref{sec:costs} and show that our batched implementations are efficient and scale to millions of input rows with a single thread. Figure~\ref{fig:micro-relational} shows the results. 

\stitle{Unary operators.} In Figure~\ref{fig:group-distinct-order}, we plot the execution time of unary operators vs the input size. 
Recall from Section~\ref{sec:costs} that \texttt{DISTINCT} and \texttt{GROUP-BY} are both based on sorting and, thus, their cost includes the cost of \texttt{ORDER-BY} for unsorted inputs of the same cardinality. To shed more light on the performance of \texttt{DISTINCT} and \texttt{GROUP-BY}, Figure~\ref{fig:group-distinct-order} only shows the execution time of their second phase, that is, after the input is sorted and, for \texttt{GROUP-BY}, before the final shuffling (which has identical performance to sorting).

For an input relation with $n$ rows, \texttt{DISTINCT} performs $n-1$ equality comparisons, one for each pair of adjacent rows. Since all these comparisons are independent, our implementation uses batching, thus, applying \texttt{DISTINCT} to the entire input in six rounds of communication (the number of rounds required for oblivious equality on pairs of 64-bit shares). As a result, \texttt{DISTINCT} scales well with the input size and can process $10M$ rows in $20s$.
\texttt{GROUP BY} is slower than \texttt{DISTINCT}, as it requires significantly more rounds of communication, linear to the input size. 
Finally, \texttt{ORDER BY} relies on our implementation of bitonic sort, where all $\frac{n}{2}$ comparisons at each level are batched within the same communication round.

\stitle{Joins.} 
The oblivious join operators in \ours hide the size of their output, thus, they compute the cartesian product between the two input relations and produce a bit share for all pairs of records, resulting in an output with $n \cdot m$ entries. We run both operators with $n=m$, for increasing input sizes, and plot the results in Figure~\ref{fig:joins}. The figure includes equi-join results for up to $100K$ rows per input and range-join results for up to $40K$ rows per input, as we capped the duration of this experiment to $5h$. 
\ours executes joins in batches without materializing their entire output at once. As a result, it can perform $10B$ equality and inequality comparisons under MPC within the experiment duration limit.

 \begin{figure}[t]
    \centering
           \begin{minipage}{.23\textwidth}
            \begin{subfigure}{\textwidth}
            \centering
  			\includegraphics[width=\linewidth]{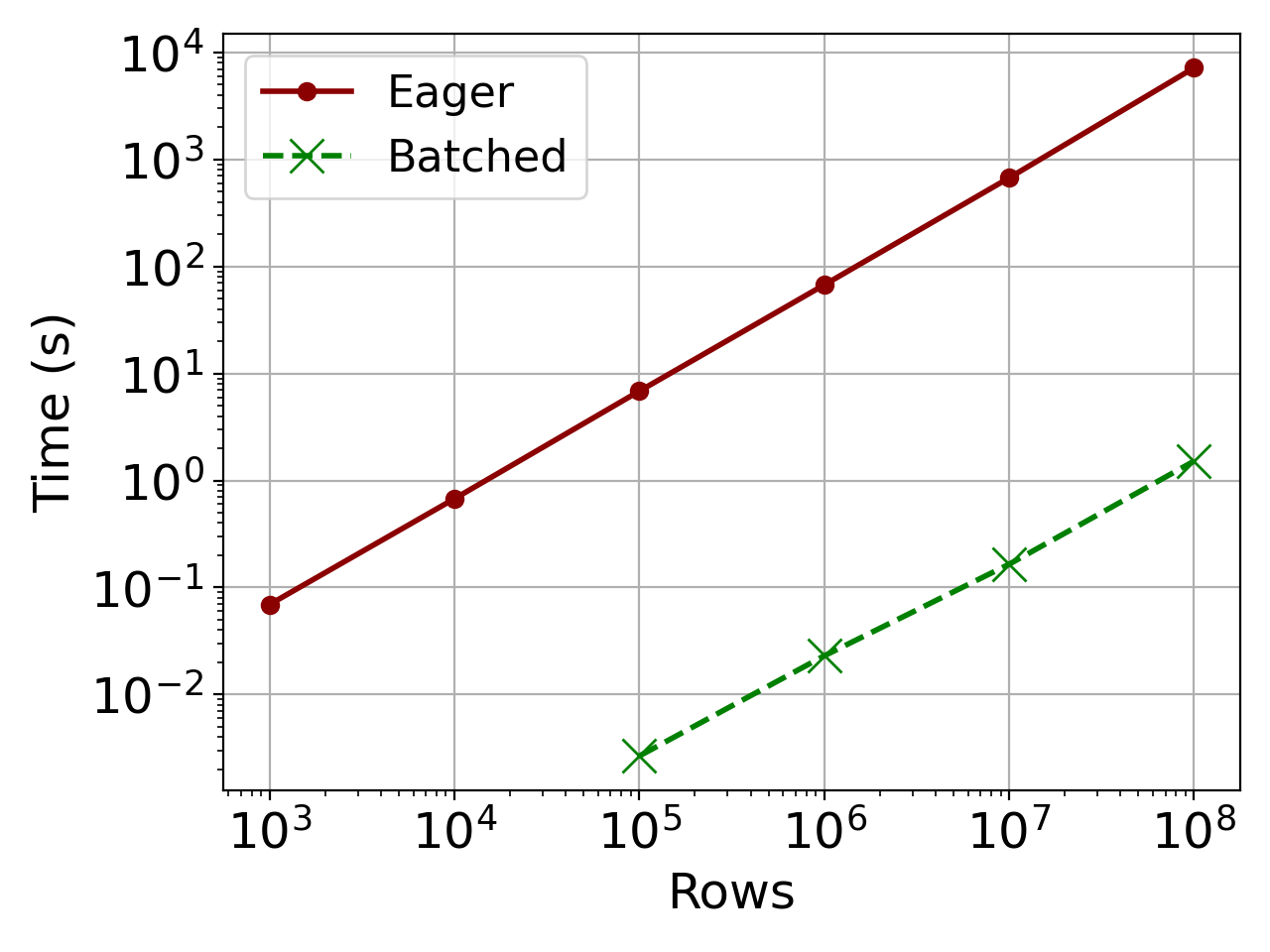}
 			 \vspace{-6mm}
 			 \caption{Eager vs. batched communication}\label{fig:micro-exchange}
            \end{subfigure}
        \end{minipage}    
        \begin{minipage}{.23\textwidth}
            \begin{subfigure}{\textwidth}
            \centering
            \includegraphics[width=\textwidth]{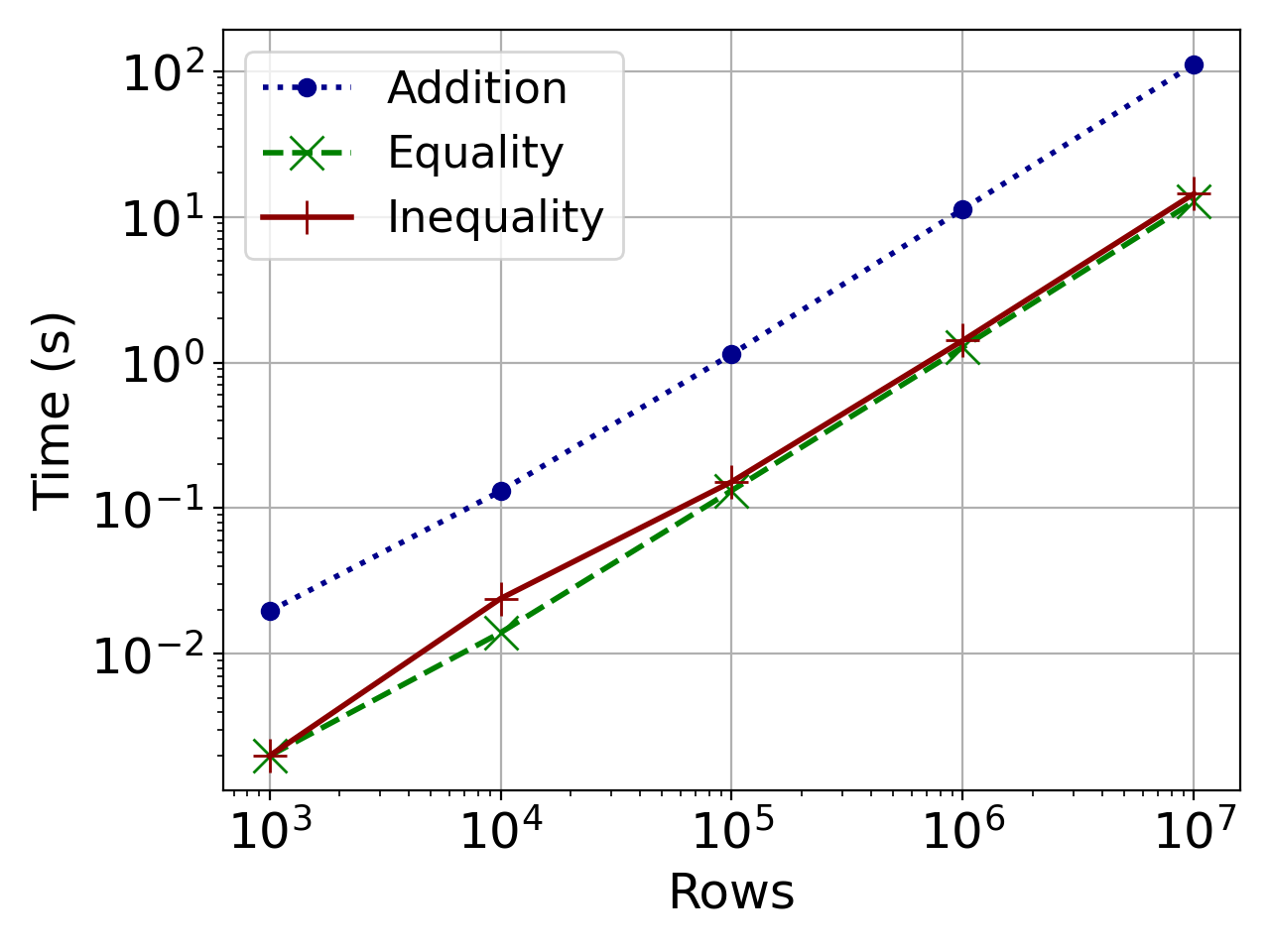}
            \vspace{-6mm}
            \caption{Comparison and addition}\label{fig:micro-comp}
            \end{subfigure}
        \end{minipage}
         \caption{Performance of oblivious primitives in \ours}\label{fig:micro-primitives} \vspace{-4mm}
    \end{figure}

\begin{table*}[t!]
\centering
\footnotesize
\begin{tabular}{cccccccc}
  \multicolumn{1}{c}{\rot{\textbf{Framework}}} &
  \multicolumn{1}{c}{\rot{\textbf{\begin{tabular}[c]{@{}c@{}}MPC Protocol\end{tabular}}}} &
  \multicolumn{1}{c}{\rot{\textbf{\begin{tabular}[c]{@{}c@{}}Information \\ Leakage\end{tabular}}}} &
  \multicolumn{1}{c}{\rot{\textbf{\begin{tabular}[c]{@{}c@{}}Trusted \\ Party\end{tabular}}}} &
  \multicolumn{1}{c}{\rot{\textbf{\begin{tabular}[c]{@{}c@{}}Query \\ Execution\end{tabular}}}} &
  \multicolumn{1}{c}{\rot{\textbf{\begin{tabular}[c]{@{}c@{}}Optimization \\ Objective\end{tabular}}}} &
  \multicolumn{1}{c}{\rot{\textbf{\begin{tabular}[c]{@{}c@{}}Optimization Conditions\end{tabular}}}}
   \\ \hline
  
\multicolumn{1}{|l|}{\begin{tabular}[c]{@{}l@{}}Conclave~\cite{Volgushev2019Conclave}\end{tabular}} &
  \multicolumn{1}{c|}{\begin{tabular}[c]{@{}c@{}}Secret Sharing /\\Garbled Circuits \end{tabular}} &
  \multicolumn{1}{c|}{\begin{tabular}[c]{@{}c@{}}Controlled \\ (Hybrid operators)\end{tabular}} &
  \multicolumn{1}{c|}{Yes} &
  \multicolumn{1}{c|}{Hybrid} &
  \multicolumn{1}{c|}{\begin{tabular}[c]{@{}c@{}}Minimize the use of\\ secure computation\end{tabular}} &
  \multicolumn{1}{c|}{\begin{tabular}[c]{@{}l@{}}1. Data owners participate in computation\\ 2. Data owners provide privacy annotations\\ 3. There exists a trusted party\end{tabular}}
  \\ \hline
  
 \multicolumn{1}{|l|}{SMCQL~\cite{Bater2017SMCQL}} &
  \multicolumn{1}{c|}{\begin{tabular}[c]{@{}c@{}}Garbled Circuits / \\ ORAM\end{tabular}} &
  \multicolumn{1}{c|}{No} &
  \multicolumn{1}{c|}{No} &
  \multicolumn{1}{c|}{Hybrid} &
  \multicolumn{1}{c|}{\begin{tabular}[c]{@{}c@{}}Minimize the use of\\ secure computation\end{tabular}} &
  \multicolumn{1}{c|}{\begin{tabular}[c]{@{}l@{}}1. Data owners participate in computation \\ 2. Data owners provide privacy annotations\end{tabular}}
  \\ \hline

  \multicolumn{1}{|l|}{Shrinkwrap~\cite{bater2018shrinkwrap}} &
  \multicolumn{1}{c|}{\begin{tabular}[c]{@{}c@{}}Garbled Circuits / \\ ORAM\end{tabular}} &
  \multicolumn{1}{c|}{\begin{tabular}[c]{@{}c@{}}Controlled \\ (Diff. Privacy)\end{tabular}} &
  \multicolumn{1}{c|}{\begin{tabular}[c]{@{}c@{}}No\end{tabular}} &
  \multicolumn{1}{c|}{Hybrid} &
  \multicolumn{1}{c|}{\begin{tabular}[c]{@{}c@{}}Calibrate padding of\\intermediate results\end{tabular}} &
  \multicolumn{1}{c|}{\begin{tabular}[c]{@{}l@{}}\hspace{0.9mm}1. Data owners participate in computation$^1$\\ \hspace{0.9mm}2. Data owners provide privacy annotations$^1$\\ \hspace{0.9mm}and intermediate result sensitivities\end{tabular}}
  \\ \hline
  
  \multicolumn{1}{|l|}{SAQE~\cite{bater2020saqe}} &
  \multicolumn{1}{c|}{Garbled Circuits} &
  \multicolumn{1}{c|}{\begin{tabular}[c]{@{}c@{}}Controlled \\ (Diff. Privacy)\end{tabular}} &
  \multicolumn{1}{c|}{\begin{tabular}[c]{@{}c@{}}No\end{tabular}} &
  \multicolumn{1}{c|}{Hybrid} &
  \multicolumn{1}{c|}{\begin{tabular}[c]{@{}c@{}}Choose sampling rate\\for approximate answers\end{tabular}} &
  \multicolumn{1}{c|}{\begin{tabular}[c]{@{}l@{}}\hspace{0.9mm}1. Data owners participate in computation$^1$\\ \hspace{0.9mm}2. Data owners provide privacy annotations$^1$\\ \hspace{0.9mm}and privacy budget \end{tabular}}
  \\ \hline
  
  \multicolumn{1}{|l|}{Senate~\cite{Poddar2021Senate} $^{2}$} &
  \multicolumn{1}{c|}{Garbled Circuits} &
  \multicolumn{1}{c|}{No} &
 \multicolumn{1}{c|}{No} &
  \multicolumn{1}{c|}{Hybrid} &
  \multicolumn{1}{c|}{\begin{tabular}[c]{@{}c@{}}Reduce joint computation\\ to subsets of parties\end{tabular}} &
  \multicolumn{1}{c|}{\begin{tabular}[c]{@{}l@{}}\hspace{0.2mm}1. Data owners participate in computation\\ \hspace{0.2mm}2. Input or intermediate relations are owned\\  \hspace{0.2mm}by subsets of the computing parties\end{tabular}}
  \\ \hline

  \multicolumn{1}{|l|}{SDB~\cite{wong2014secure, he2015sdb} $^{3}$} &
  \multicolumn{1}{c|}{\begin{tabular}[c]{@{}c@{}}Secret Sharing\end{tabular}} &
  \multicolumn{1}{c|}{\begin{tabular}[c]{@{}c@{}}Yes \\ (operator dependent)\end{tabular}} &
  \multicolumn{1}{c|}{No} &
  \multicolumn{1}{c|}{Hybrid} &
  \multicolumn{1}{c|}{\begin{tabular}[c]{@{}c@{}} Reduce data encryption\\ and decryption costs\end{tabular}} &
  \multicolumn{1}{c|}{\begin{tabular}[c]{@{}l@{}}\hspace{-0.2mm}1. Data owner participates in computation\\ \hspace{-0.2mm}2. Data owner provides privacy annotations\end{tabular}}
  \\  \hline
  
  \\ \hline
  \multicolumn{1}{|l|}{\textbf{\ours}} &
  \multicolumn{1}{c|}{\begin{tabular}[c]{@{}c@{}}Repl. Secret Sharing\end{tabular}} & 
  \multicolumn{1}{c|}{No} &
  \multicolumn{1}{c|}{No} &
  \multicolumn{1}{c|}{\begin{tabular}[c]{@{}c@{}}End-to-end \\ under MPC\end{tabular}} &
  \multicolumn{1}{c|}{\begin{tabular}[c]{@{}c@{}} Reduce MPC costs\\ (Section~\ref{sec:optimizations})\end{tabular}} &
  \multicolumn{1}{c|}{None}
  \\ \hline
\end{tabular}
\begin{flushleft}
\hspace{2mm}\emph{$^{1}$ Shrinkwrap and SAQE build on top of SMCQL's information flow analysis and inherit its optimizations along with their conditions.}\\
\hspace{2mm}\emph{$^{2}$ Senate provides security against malicious parties whereas all other systems adopt a semi-honest model.}\\
\hspace{2mm}\emph{$^{3}$ SDB adopts a typical DBaaS model with one data owner and does not support collaborative analytics.} \\
\end{flushleft}
\caption{Summary of MPC-based software solutions for relational analytics. Hybrid query execution is  feasible when data owners participate in the computation. The rest of the optimizations supported by each system are applicable under one or more of the listed conditions.}\label{tbl:comparison}
\end{table*}

\subsection{Performance of \ours's primitives}\label{sec:exp-micro}
To better understand the results of the previous sections, we now use a set of micro-benchmarks and evaluate the performance of \ours's MPC primitives in \texttt{AWS-LAN}.

\stitle{Effect of message batching on communication latency.}
In the first experiment, we measure the latency of inter-party communication using two messaging strategies.
Recall that, during a message exchange, each party sends one message to its successor and receives one message from its predecessor on the `ring'.
\emph{Eager} exchanges data among parties as soon as they are generated, thus,
producing a large number of small messages. The \emph{Batched} strategy, on the other hand,
collects data into batches and exchanges them only when computation cannot otherwise make progress, thus,
producing as few as possible, albeit large messages. 

We run this experiment with increasing data sizes and measure the total time from initiating the exchange
until all parties complete the exchange. Figure~\ref{fig:micro-exchange} shows the results. We see that batching provides two to four orders of magnitude lower latency than eager
messaging. Using batching in our experimental setup, parties can exchange $100M$ 64-bit data shares in $2s$.
These results reflect the network performance in our cloud testbed. We expect better performance in dedicated clusters with high-speed networks and higher latencies if the computing parties communicate over the internet.

\stitle{Performance of secure computation primitives.}
We now evaluate the performance of oblivious primitives that require communication among parties. These include equality, inequality,  and addition with the ripple-carry adder.
In Figure~\ref{fig:micro-comp} we show the execution time of oblivious primitives as we increase the input size from $1K$ rows  to $10M$ rows. All primitives scale well with the input size as they all depend on a constant number of communication rounds. Equality
requires six rounds. Inequality requires seven rounds and more memory than equality. Boolean addition is not as memory- and computation-intensive as inequality, but requires a higher number of rounds (64). 

\section{Related Work}\label{sec:related}

\stitle{Relational MPC frameworks.} We distinguish two main lines of work in this space that are often combined, as shown in Table~\ref{tbl:comparison}. Hybrid query execution~\cite{Bater2017SMCQL,Volgushev2019Conclave,DBLP:conf/ndss/ChowLS09,secure-db-services} improves performance by splitting the query plan into a plaintext part and an oblivious part. The second line of efforts includes frameworks that trade off secure query performance with controlled information leakage~\cite{bater2018shrinkwrap,bater2020saqe,Volgushev2019Conclave,wong2014secure, he2015sdb}. 
More recently, Senate~\cite{Poddar2021Senate} combined hybrid execution with a technique that reduces secure computation to subsets of the computing parties. 
All optimizations proposed in these works are applicable under certain conditions on data sensitivity, input ownership, and the role of data owners in the computation. For example, minimizing the use of MPC via hybrid execution is only feasible when data owners can compute part of the query locally on their plaintext data. SMCQL~\cite{Bater2017SMCQL}, SDB~\cite{wong2014secure}, and Conclave~\cite{Volgushev2019Conclave} can further sidestep MPC when attributes are annotated as non-sensitive, Shrinkwrap~\cite{bater2018shrinkwrap} and SAQE~\cite{bater2020saqe} calibrate leakage based on user-provided privacy budgets, and Senate reduces joint computation when some relations are owned by subsets of the computing~parties. This is pretty common in peer-to-peer MPC but does not occur in a typical outsourced setting where all computing parties have shares of~the~data. 

Our approach has several advantages over, and is also complementary with, many of these prior techniques. \ours's optimizations are agnostic to data ownership and retain the full security guarantees of MPC, merely optimizing its execution. In settings where prior works apply, our optimizations can be incorporated into existing systems to further optimize the oblivious query part.

\stitle{Oblivious operators and algorithms.} Various related works focus on standalone oblivious relational operators, e.g.
building group-by from oblivious sort~\cite{DBLP:journals/iacr/JonssonKU11},
building equi-joins~\cite{Krastnikov2020Efficient, database-info-sharing,psi-cuckoo-hashing, DBLP:conf/ccs/MohasselRR20},
or calculating common aggregation operators~\cite{privacy-preserving-query}. 
Our research is driven by real-world applications that typically require oblivious evaluation of queries with multiple operators. 
Recently, Wang et al. \cite{10.1145/3448016.3452808} presented a secure version of the Yannakakis' algorithm, while Ion et al.~\cite{DBLP:journals/iacr/IonKNPRSSSY19} and Buddhavarapu et al.~\cite{cryptoeprint:2020:599} studied unique-key joins followed by simple aggregations. 
None of these works provides general cost-based MPC query optimization and they all operate in the peer-to-peer setting. However, any techniques that can be adapted for the outsourced setting could be incorporated in~\ours.

\stitle{Enclave-based approaches.}
In this line of work, parties process the actual data within a physically protected environment. 
Enclave-based approaches aim to minimize RAM requirements, pad intermediate results, and hide access patterns when accessing untrusted storage.
The works by Agrawal et al.~\cite{sovereign-joins} and Arasu et al.~\cite{ArasuK2014Oblivious} focus on database queries in this setting.
ObliDB~\cite{eskandarian2019oblidb}, Opaque~\cite{zheng2017opaque}, and StealthDB~\cite{stealthdb} rely on secure hardware (Intel's SGX). 
OCQ~\cite{DBLP:conf/eurosys/DaveLPGS20} builds on Opaque and introduces additional optimizations that reduce intermediate result padding by leveraging FK constraints between private and public relations.
Enclave-based systems typically achieve better performance than MPC-based systems but require different trust assumptions (as an alternative to cryptography) and are susceptible to  attacks~\cite{sgx-attack3, sgx-attack1, sgx-attack2, sgx-attack5,sgx-attack4, DBLP:conf/uss/BulckMWGKPSWYS18, sgx-attack6, DBLP:conf/uss/LeeJFTP20}.

\stitle{Encrypted databases.} 
Existing practical solutions in secure database outsourcing~\cite{DBLP:conf/sp/FullerVYSHGSMC17} avoid the need for a non-collusion assumption, but they reveal or ``leak''
information to the single database server.
Systems based on property-based encryption
like CryptDB \cite{cryptdb} offer full SQL support and legacy compliance, but each query reveals information  that can be used in reconstruction attacks~\cite{kellaris,DBLP:conf/ccs/NaveedKW15,DBLP:conf/ccs/GrubbsLMP18}.
Systems based on structural encryption~\cite{structural-encryption, arx, blindseer, DBLP:conf/ndss/CashJJJKRS14, kafedb} 
provide semantic security for data at rest and better protection, but do not eliminate access pattern~leaks. 
SDB~\cite{wong2014secure, he2015sdb} uses secret-sharing in the typical client-server model but its protocol leaks information to the database server. 
Finally, Cipherbase~\cite{cypherbase} is a database system that relies on a secure coprocessor. 
These systems only support one data owner, and it would require considerable performance overhead to extend to our setting due to the need for public key encryption to support queries that span multiple datasets \cite{BoschHJP14}. 
 
\stitle{Differential privacy.} Systems like 
DJoin \cite{djoin},
DStress \cite{dstress},
and the work of He et al. \cite{DBLP:conf/ccs/HeMFS17}
use the concept of differential privacy to ensure that the output of a query reveals little about any one input record.
This property is independent of (yet symbiotic with) MPC's security guarantee that the act of computing the query reveals no more than what may be inferred from its output. \ours could be augmented to provide differentially private outputs if desired. 

Shrinkwrap \cite{bater2018shrinkwrap} and SAQE \cite{bater2020saqe} achieve better efficiency by relaxing security for the computing parties only up to differentially private leakage.
This is effectively the same guarantee as above when the computing and result parties are identical, but is weaker when they are different.
\ours does not leak any information. 

\stitle{ORAM-based approaches.} Oblivious RAM~\cite{oram1,oram2} allows for compiling arbitrary programs into oblivious ones by carefully distorting access patterns to eliminate leaks. 
ORAM-based systems like SMCQL \cite{Bater2017SMCQL} and Obladi~\cite{obladi} hide access patterns but the flexibility of ORAM comes at high cost to throughput and latency.
Two-server distributed ORAM systems like Floram \cite{DBLP:conf/ccs/DoernerS17} and SisoSPIR \cite{DBLP:conf/ctrsa/IshaiKLO16} are faster but require the same non-collusion assumption as in this work. \ours does not rely on ORAM; instead, we implement specific database operators with a data-independent control~flow.

\stitle{FHE-based approaches.} Fully Homomorphic Encryption (FHE) protocols~\cite{fhe} allow arbitrary computations directly on encrypted data with strong security guarantees. Although many implementations exist~\cite{fhe-impl,fhe-impl2,fhe-impl3,fhe-impl4,fhe-impl5,fhe-impl6}, this approach is still too computationally expensive for the applications we consider in this work.

\stitle{Other MPC frameworks.} The recent advances in MPC have given rise to many practical 
general-purpose MPC frameworks like ABY \cite{aby}, ABY3~\cite{aby3}, Jiff~\cite{jiff}, SCALE-MAMBA \cite{scale-mamba}, ObliVM\cite{oblivm}, Obliv-C~\cite{obliv-c}, and ShareMind~\cite{sharemind}; we refer readers to Hastings et al.~\cite{DBLP:conf/sp/HastingsHNZ19} for an overview of these frameworks.
Some of these frameworks support standalone database operators (e.g.~\cite{sharemind, aby3}) but do not address query composition and optimization under MPC. Splinter~\cite{splinter} uses function secret sharing to protect private queries on public data, which is exactly the opposite to our setting. This system supports a subclass of SQL queries that do not include private joins.

\section{What's next?}\label{sec:next}

We see several exciting research directions for the crypto, systems, and data management communities:

\stitle{MPC query optimizers.}
The optimizations in \S~\ref{sec:optimizations} are by no means exhaustive and there are many opportunities for continued research in this space. For example, one could consider alternative oblivious operators with different tradeoffs or operators that leverage public information about the data schema. Most importantly, our MPC cost analysis has revealed that optimal plans in a plaintext evaluation are not necessarily optimal under MPC (and vice versa) and   
there is a need for robust MPC query optimizers that take into account the characteristics of secure computation.
Recently, Senate~\cite{Poddar2021Senate} did some nice work in this direction but the proposed optimizations are not applicable in a typical outsourced setting where all computing parties have shares of the data. 
The logical query optimizations supported by \ours are general and can serve as the foundation to design an efficient optimization framework that maintains the security guarantees (e.g., semi-honest or malicious, with the same adversarial threshold) of the underlying MPC protocol. 

\stitle{Parallelism and oblivious hashing.}
Task and data parallelism offer the potential for improved performance and scalability. Extending oblivious operators to work in a task-parallel fashion is straight-forward (e.g. for bitonic sort) but data-parallel execution requires additional care. In a plaintext data-parallel computation, data are often partitioned using hashing: the data owners agree on a hash function and hash the input records into buckets, so that subsequent join and group-by operations only need to compare records within the same bucket.
In MPC, data parallelism can be achieved via oblivious hashing, with care taken to ensure that the bucket sizes do not reveal the data distribution or access patterns. Indeed, many private set intersection algorithms leverage this technique in a setting where the input and computing parties are identical \cite{DBLP:conf/uss/Pinkas0SZ15}. To achieve better load balancing of keys across buckets and keep the bucket size low, one can use Cuckoo hashing, as in \cite{psi-cuckoo-hashing}.
It is an interesting direction to design oblivious hashing techniques in the outsourced setting, where  
data owners generate and distribute secret shares along with their corresponding bucket IDs to reduce the cost of oblivious join and group-by~operators.

\stitle{System optimizations and HW acceleration.} 
Our experience in developing new implementations of MPC protocols has been very valuable in revealing
performance trade-offs in secure computation. We believe that several database and systems optimization techniques will prove useful in improving MPC performance. 
For example, to hide intermediate results and access patterns, some oblivious operators append a new column of secret shares to their input (e.g., selection),  while others shuffle entire rows (e.g., order-by).  Exploring the performance trade-offs between column- and row-oriented data representation is an interesting direction. Significant performance improvement could also be achieved by hiding the overhead of repeated local computations required for security, such as random number generation. Our preliminary analysis suggests that, in certain cases, random number generation accounts for up to $70\%$ of the total query execution time.
Employing OS-level optimizations in the spirit of MAGE~\cite{Kumar2021MAGE} is another promising direction. 
Lastly, while \ours takes a software-only approach, one could also offload certain MPC primitives to hardware accelerators (e.g., ~\cite{FangIL17, FrederiksenJN14, SonghoriZD0SK16, HussainRGK18}) to further improve throughput and latency.

\stitle{Malicious security.}
While the current work focuses on semi-honest security, it provides a strong foundation for achieving malicious security in the future. \ours protects data using the replicated secret sharing scheme of Araki et al. \cite{ArakiFLNO16}, which can be extended to provide malicious security with low computational cost \cite{billion-gates-mpc}.
By optimizing MPC rather than sidestepping it, our approach has an advantage over prior work \cite{Poddar2021Senate}: we do not need to take additional non-trivial measures to protect the integrity of local pre-processing steps.

\section*{Acknowledgments}

The authors are grateful to Kinan Dak Albab, Azer Bestavros, and Ben Getchell for their valuable feedback, and to the Mass Open Cloud for providing access to their cloud for initial experiments. We would also like to thank Eric Chen for his help with automating the deployment of Secrecy to multiple Clouds. The fourth author's work is supported by the DARPA SIEVE program under Agreement No.\ HR00112020021 and the National Science Foundation under Grants No.~1414119, 1718135, 1801564, and 1931714.


\bibliographystyle{ACM-Reference-Format}
\bibliography{refs}


\end{document}